\documentclass[useAMS,usenatbib]{mn2e} \usepackage{amsmath,amssymb}
\usepackage{graphicx,mathptm} \usepackage{times} \usepackage{natbib}
\usepackage{psfig}
\newcommand{\kms}{$\mbox{km/s}$\,}
\newcommand{\Mpch}{$h^{-1}\,\mbox{Mpc}$\,}
\newcommand{\xii}{$\xi(r_\perp,r_\parallel)$\,}
\newcommand{\xiiz}{$\xi(s_\perp,s_\parallel)$\,}
\newcommand{\Om}{$\Omega_{\rm M}$\,}

\title[Disentangling dynamics and geometry] {Cosmology with clustering
  anisotropies: disentangling dynamic and geometric distortions in
  galaxy redshift surveys}

\author[Marulli et al.]  {Federico Marulli$^{1,2,3}$, Davide
  Bianchi$^{4,5}$, Enzo Branchini$^{5,6,7}$, Luigi Guzzo$^5$,
  \newauthor Lauro Moscardini$^{1,2,3}$ and Raul
  E. Angulo$^8$\\ $^1$Dipartimento di Astronomia, Alma Mater Studiorum
  - Universit\`a di Bologna, via Ranzani 1, I-40127 Bologna,
  Italy\\ $^2$INAF, Osservatorio Astronomico di Bologna, via Ranzani
  1, I-40127 Bologna, Italy\\ $^3$INFN, Sezione di Bologna, viale
  Berti Pichat 6/2, I-40127 Bologna, Italy\\ $^4$Dipartimento di
  Fisica, Universit\`a degli Studi di Milano, via Celoria 16, I-20133
  Milano, Italy\\ $^5$INAF, Osservatorio Astronomico di Brera, via
  Brera 28, I-20122 Milano, via E. Bianchi 46, I-23807 Merate, Italy
  \\ $^6$Dipartimento di Fisica, Universit\`a degli Studi ``Roma
  Tre'', via della Vasca Navale 84, I-00146 Roma, Italy\\ $^7$INFN,
  Sezione di Roma Tre, via della Vasca Navale 84, I-00146 Roma,
  Italy\\ $^8$Max-Planck Institut F\"ur Astrophysik,
  Karl-Schwarzshild-Stra{\ss}e 1, 85740 Garching, Germany\\ }

\begin{document}

\maketitle

\label{firstpage}


\begin{abstract}
We investigate the impact of different observational effects affecting
a precise and accurate measurement of the growth rate of fluctuations
from the anisotropy of clustering in galaxy redshift surveys. We focus
here on redshift measurement errors, on the reconstruction of the
underlying real-space clustering and, most importantly, on the
apparent degeneracy existing with the geometrical distortions induced
by the cosmology-dependent conversion of redshifts into distances.  We
use a suite of mock catalogues extracted from large N-body
simulations, focusing on the analysis of intermediate, mildly
non-linear scales ($r<50 $\Mpch) and apply the standard ``dispersion
model'' to fit the anisotropy of the observed correlation function
\xii.  We first verify that redshift errors up to $\delta z\sim0.2\%$
(i.e. $\sigma_z\sim 0.002$ at $z=1$) have a negligible impact on the
precision with which the specific growth rate $\beta$ can be measured.
Larger redshift errors introduce a positive systematic error, which
can be alleviated by adopting a Gaussian distribution function of
pairwise velocities.  This is, in any case, smaller than the
systematic error of up to $10\%$ due to the limitations of the
dispersion model, which is studied in a separate paper.  We then show
that $50\%$ of the statistical error budget on $\beta$ depends on the
deprojection procedure through which the real-space correlation
function, needed for the modelling process, is obtained. Finally, we
demonstrate that the degeneracy with geometric distortions can in fact
be circumvented.  This is obtained through a modified version of the
Alcock-Paczynski test in redshift-space, which successfully recovers
the correct cosmology by searching for the solution that optimizes the
description of dynamical redshift distortions. For a flat cosmology,
we obtain largely independent, robust constraints on $\beta$ and on
the mass density parameter, \Om.  In a volume of $2.4
(h^{-1}\,\mbox{Gpc})^3$, the correct \Om is obtained with $\sim12\%$
error and negligible bias, once the real-space correlation function is
properly reconstructed.
\end{abstract}



\section {Introduction}

The large scale structure of the Universe is one of the main
observational probes to discriminate among competing cosmological
models and estimate their fundamental parameters, some related to
space-time geometry, some related to the density fluctuations.  The
growth rate of density fluctuations, $f(z)$, belongs to the second
category.  Since the pioneering works of \citet{kaiser1987} and
\citet{hamilton1998}, it was clear that one of the most promising ways
to determine $f(z)$ is to exploit the apparent anisotropy in the
clustering of galaxies induced by peculiar velocities, an effect
commonly known as redshift-space distortions (RSD).

Being $f(z)$ directly sensitive to the mean density of matter, for
some time RSD have been used to estimate the mass density parameter
\Om \citep[e.g.][]{peacock2001, hawkins2003, daangela2005a,
  daangela2005b, ross2007, ivashchenko2010}.  Later on, with the
advent of other, more precise methods to estimate \Om, like the
Barionic Acoustic Oscillations (BAO), RSD began to be considered as a
sort of ``noise'' to be marginalized over \citep{seo2003, seo2007}.
New interest on RSD arose when it was realized that, if not
marginalized over, they could tighten constraints over cosmological
parameters \citep{amendola2005}, and in particular when it was shown
\citep{guzzo2008, zhang2008} that they could represent a formidable
tool to discriminate between a dark energy (DE) scenario and a modified
gravity theory for the origin of cosmic acceleration.  A number of
forecast papers followed rapidly \citep[e.g.][]{linder2008, wang2008,
  song2009}, as well as applications to existing and new datasets
\citep{cabre2009a, cabre2009b, blake2011a}. As such, RSD are now
regarded as one of the most promising techniques to extract precise
estimates of $f(z)$ from future redshift surveys. Besides their use as
a probe to constrain alternative gravity theories, RSD can also be
exploited in many different contexts. For instance, it has been
recently demonstrated that RSD robustly constrain the mass of relic
cosmological neutrinos \citep{marulli2011} and could be used to detect
interactions in the dark sector \citep{marulli2012}. Moreover, they
can be of some help in astrophysical contexts, e.g. to investigate the
dynamical properties of the warm-hot intergalactic medium
\citep{ursino2011}.

To effectively discriminate among competing cosmological models,
however, one needs to measure the growth rate with per cent accuracy.
This goal has prompted several works aimed at better identifying and
characterizing the sources or uncertainty.  One aspect of the problem
is how to optimally infer the growth rate of fluctuations from the
measured RSD quantities.  The anisotropic pattern of galaxy redshifts
in 3D space allows one to estimate the distortion parameter
$\beta(z)\equiv f(z)/b(z)$.  Then, to obtain $f(z)$, one needs in
principle an independent estimate of the galaxy bias parameter $b(z)$
which is ill-constrained by theory and difficult to measure from the
data.  For this reason, it has been suggested \citep[see
  e.g.][]{white2009, percival2009, song2009} to express the
constraints in terms of the observed product $\beta(z)
\sigma^{gal}_8(z)$, which in the linear bias hypothesis equals the
theoretical combination $f(z)\sigma_8(z)$.  In these expressions
$\sigma^{gal}_8$ and $\sigma_8$ are the rms values in spheres of 8
\Mpch of respectively the galaxy counts and the mass.  In this way,
the values of $\beta(z) \sigma^{gal}_8(z)$ obtained from the data at
different redshifts can be self-consistently compared to the curves
$f(z)\sigma_8(z)$ predicted by the theory (once they are nevertheless
normalized to a reference $\sigma_8$ as provided, e.g., by CMB
measurements).

Another aspect of the problem is the impact of the observational
errors, related to redshift measurements and to the nature of the
objects used to trace the mass inhomogeneities.  This is commonly
addressed using the Fisher information matrix \citep{fisher1935,
  tegmark1997}, which has become the standard method to forecast
statistical errors in the cosmological parameters expected from
planned redshift surveys \citep[see e.g.][]{sapone2007, wang2008,
  linder2008, white2009, percival2009, wang2010, acquaviva2010,
  fedeli2011,carbone2011a, carbone2011b, carbone2012, diporto2012,
  diporto2012b, majerotto2012}. One limitation in the use of the
Fisher matrix is that it relies on assumptions (e.g. the Gaussian
nature of errors) that in some applications might not be fully
justified.  In addition, the Fisher matrix formalism cannot say
anything about systematic errors that may dominate the error budget in
future experiments aiming at per cent accuracy.  Finally, the use of
the Fisher matrix is justified on large scales, where linear theory
applies, but may fail on smaller scales because of the improper
treatment of deviations from linearity in dynamics and bias.  In all
these cases one needs to use numerical simulations.

Early attempts to test the accuracy of the linear model of RSD
\citep{kaiser1987} through numerical simulations and develop improved
corrections \citep{scoccimarro2004, tinker2006} have been recently
extended, following the renewed interest on this topic
\citep{taruya2011, jennings2011b, jennings2011a, okumura2011,
  kwan2012, samushia2011}.  The work presented here is part of this
ongoing effort to understand the limitations of RSD estimators and
bring this technique at the level required by precision cosmology.

Finally, a potentially relevant source of systematic error in the
measurement of RSD is related to intrinsic uncertainty on the
background cosmology, which is needed when converting redshifts into
distances.  This introduces further {\sl geometric distortions} (GD)
in the observed clustering pattern, which in principle can be used to
estimate the background cosmological parameters through the so-called
Alcock-Paczynski (AP) test \citep{alcock1979}.  Several methods and
attempts to implement the AP test using different datasets have been
proposed \citep{alcock1979, phillipps1994, ryden1995, ryden1996,
  marinoni2010}, including those that look for anisotropies in 3D
clustering \citep{ballinger1996, matsubara1996, popowski1998, hui1999,
  matsubara2000, mcdonald2003, nusser2005, barkana2006, kim2007,
  padmanabhan2008}.  In practice, GD have a smaller amplitude than and
are degenerate with RSD \citep{simpson2010}, such that only with very
high quality data one can hope to disentangle the two effects. Early
applications of the AP test thus failed to get significant results
\citep{outram2004, hoyle2002, daangela2005a, daangela2005b, ross2007},
but the situation is rapidly improving \citep{chuang2012,
  blake2011a}. It is fairly clear that GD will play a fundamental role
in the estimate of cosmological parameters from future surveys
\citep{seo2003, simpson2010, kazin2012, taruya2011, samushia2012,
  hawken2012}, but also that, if not properly accounted for, they
represent one further source of systematic error in the measurement of
$f(z)$ from RSD.

In this paper we explore how the precision and accuracy of $f$ depend
on the limitations of i) using observed quantities (angles and
redshift) to infer cosmological distances in redshift-space, and ii)
using them to construct and model two-point statistics.  We therefore
first assess the impact on the recovered $\beta$ of redshift
measurement errors and of the intrinsic uncertainty on the real-space
two-point correlation function due to the deprojection procedure. We
then look in detail into the effect of GD and introduce a simple
technique to isolate RSD.  We show that, under the hypothesis of a
flat background, this allows us to simultaneously measure $\beta$ and
$\Omega_{\rm M}$.  Finally, we relax the flat background assumption
and investigate how GD can be used to constrain the cosmological
parameters that enter the Hubble function. These tests focus on
intermediate scales ($r<50$\Mpch), where non-linear effects cannot be
neglected. These are the scales where RSD have the highest
signal-to-noise, also in last-generation surveys like WiggleZ
\citep{blake2011a}, VIMOS Public Extragalactic Redshift Survey
\citep[VIPERS]{guzzo2012} and SDSS-III BOSS \citep{eisenstein2011}.

In a parallel, complementary work \citep{bianchi2012}, we use the same
numerical methods adopted here to study the dependence of the
uncertainty on the measured growth rate on typical survey parameters,
as the volume and the density and bias of the adopted tracers.

The paper is organised as follows. In \S \ref{sec:tools} we start
describing the exploited set of N-body simulations and how to
construct mock halo catalogues. Then we give a general overview on RSD
and GD.  In \S \ref{sec:rsd} we investigate the impact of redshift
errors, deprojection effects and geometric uncertainties on the
estimate of the RSD parameters. The method to disentangle GD and RSD
is analysed in \S \ref{sec:AP}. Finally, in \S \ref{sec:conclusions},
we draw our conclusions.


\section {Theoretical tools}
\label{sec:tools}

In this section we review the theoretical tools we use to derive the
random and systematic errors in the estimate of $f(z)$ from RSD.
Since we are not concerned on galaxy bias, we will focus on the
distortion parameter $\beta$.


\subsection {N-body simulations and mock halo catalogues}

Since we are interested in RSD on intermediate scales, our analysis
will rely on mock catalogues extracted from N-body simulations that
fully account for non-linear dynamics.  The need for a sufficiently
large number of independent mock catalogues, each one with a volume
matching that of the typical ongoing redshift surveys, imposes to
consider very large computational boxes.

The `Baryon Acoustic Simulation at the ICC', {\small BASICC}, meets
these requirements \citep{angulo2008}.  This simulation follows the
dynamical evolution of $1448^3$ dark matter (DM) particles with mass
$M_{\rm part}=5.49\times10^{10}\,h^{-1}{\rm M}_{\odot}$ in a box of
$1340$ \Mpch, using a memory-efficient version of the {\small
  GADGET-2} code \citep{springel2005gadget2}.  The cosmological model
adopted is a $\Lambda $CDM universe with $\Omega_{\rm M}=0.25$,
$\Omega_{\rm \Lambda}=0.75$, $h\equiv H_0/100\, {\rm km\, s^{-1}
  Mpc^{-1}}=0.73$, $\sigma_8=0.9$. A detailed description of this
simulation can be found in \citet{angulo2008}.  DM haloes have been
identified by linking more than $20$ particles with a standard
friends-of-friends algorithm, so that the minimum halo mass is $M_{\rm
  min}=20\cdot M_{\rm part}\simeq 1.1\cdot10^{12} h^{-1}{\rm
  M}_{\odot}$.

Since we are not interested in realistic models of galaxies, we simply
identify mock galaxies in the simulation with the DM haloes with
$M>M_{\rm min}$.  Their {\em effective bias} is
\begin{equation}
b_{\rm eff}(z) = \frac{\int_{M_{\rm min}}^{\infty} n(M,z)
  b(M,z)dM}{\int_{M_{\rm min}}^{\infty} n(M,z)dM} \; ,
\label{eq:bias} 
\end{equation}
where $n(M,z)$ is the halo mass function and $b(M,z)$ is their linear
bias.  Our goal is to assess the errors on $\beta$, so we need a
reference value to compare with. This is obtained by dividing the
expected value $f(z)=\Omega_{\rm M}^{0.545}(z)$ by the effective bias
obtained from Eq.~(\ref{eq:bias}).  As we have seen in the companion
paper \citep{bianchi2012}, the effective bias of our mock DM haloes is
in good agreement with the model of \citet{tinker2010}, that we will
consider as our reference bias model. For comparison, we will also
consider the bias function predicted by \citet{sheth_mo_tormen2001}.
The $\sim 10\%$ discrepancy between these two models gives an idea of
current theoretical uncertainties.

Because of the limited mass resolution of the simulation, we are
forced to ignore substructures within haloes.  As a result, the
largest haloes should represent cluster of galaxies collapsed into
single objects rather than individual galaxies.  This limitation has
the effect of underestimating the contribution of small scale pairwise
velocity dispersion to RSD.  We can think of this undesired effect as
an attempt of mimicking observational constraints, like the fact that
fiber collision imposes a limit to the number of spectra that can be
taken in crowded areas, or catalogue-making procedures like that of
collapsing clusters or groups of galaxies into a single object.
Further analyses of mock surveys from the Millennium run demonstrate
that the conclusions of our study do not depend on the missing
sub-structure in the BASICC DM halo catalogues \citep{bianchi2012}.

Moreover, since we want to mimic redshift surveys designed to explore
the Universe at the epoch in which the accelerated expansion has
started, we will mainly focus on the simulation output at $z=1$.
Outputs at $z=0.5$ and $z=2$ have only been considered to check the
robustness of our results.  To construct the mock halo catalogues, we
consider a local ($z=0$) observer and place the centre of the $z=1$
snapshot at the corresponding comoving distance $D_{\rm C}(z=1)$,
  where
\begin{equation}
D_{\rm C}(z)=c\int_0^z\frac{dz_c'}{H(z_c')} \; ,
\label{eq:distance}
\end{equation} 
and the Hubble expansion rate is:
\begin{multline}
H(z) = H_0\left[\Omega_{\rm M}(1+z)^3+\Omega_{\rm
    k}(1+z)^2\right. \\
  \left. +\Omega_{\rm DE}\exp\left(3\int_0^z\frac{1+w(z)}{1+z}\right)\right]^{0.5} \; ,
\label{eq:Hubble}
\end{multline}
$\Omega_{\rm k}=1-\Omega_{\rm M}-\Omega_{\rm DE}$, $w(z)$ is the DE
equation of state, and the contribution of radiation is assumed
negligible.

In doing so, we neglect structure evolution within the box.  Unless
otherwise specified, our error analysis will be performed using 27
independent mock catalogues obtained by dividing the computational box
in $3^3$ subcubes, each one about twice the size of the VIPERS survey.
The mean comoving source density in our mocks is $0.003
(h/\mbox{Mpc})^3$.

To specify the distribution of DM haloes in redshift-space, we take
the comoving coordinates of each object and compute its angular
position and observed redshift:
\begin{equation}
z_{\rm obs}=z_{\rm c}+\frac{v_\parallel}{c}(1+z_{\rm c}) + \frac{\sigma_v}{c} \; ,
\label{eq:redshift}
\end{equation}
where $z_{\rm c}$ is the {\em cosmological} redshift due to the Hubble
recession velocity at the comoving distance of the halo, $v_\parallel$
is the line-of-sight component of its center of mass velocity, $c$ is
the speed of light and $\sigma_v$ is the random error in the measured
redshift (expressed in \kms).


\subsection{Modelling Redshift Distortions}
\label{sec:mod}
RSD are induced by galaxy peculiar velocities.  On large scales, where
peculiar motions are coherent, RSD will be different than on small
scales, where the velocity field is dominated by incoherent motions
within virialized structures. An effective way to characterize these
distortions is by mean of two-point statistics, like the power
spectrum or the two-point correlation function. In this work we focus
on the latter. We will refer to the redshift-space spatial coordinates
using the vector $\vec{s}$, whereas we will use $\vec{r}$ to indicate
the real-space ones.

The estimate of the spatial two-point correlation function $\xi(r)$ is
based on counting galaxy pairs separated by a relative distance
$\vec{r}$.  To characterize RSD it is convenient to decompose the
distances in two components perpendicular and parallel to the
line-of-sight, $\vec{r}=(r_\perp,r_\parallel)$.  In absence of
peculiar velocities, the iso-correlation contours of \xii are circles
in the $(r_\perp,r_\parallel)$ plane. RSD modify these contours in a
characteristic fashion: for large values of $r_\perp$ coherent motions
squash the contours along the perpendicular direction, whereas for
small $r_\perp$ incoherent motions elongate the contours along the
parallel direction, generating the so-called ``fingers of God" effect
\citep{jackson1972}.

In the linear regime, the velocity field can be determined directly
from the density field and the RSD amplitude is proportional to
$\beta$.  In this limit and in the distant observer approximation, the
two-point correlation function in redshift-space can be written in the
compact form:
\begin{equation} 
\xi(s_\perp, s_\parallel)_{\rm lin} =
\xi_0(s)P_0(\mu)+\xi_2(s)P_2(\mu)+\xi_4(s)P_4(\mu) \; ,
\label{eq:ximodellin}
\end{equation}
where $\mu=\cos\theta=s_\parallel/s$ is the cosine of the angle
between the separation vector and the line of sight,
$s=\sqrt{s_\perp^2+s_\parallel^2}$ and $P_l$ are the Legendre
polynomials \citep{kaiser1987, lilje1989, mcgill1990, hamilton1992,
  fisher1994}. The multipoles of \xiiz can be written as follows:
\begin{equation} 
\xi_0 = \left(1+ \frac{2\beta}{3} + \frac{\beta^2}{5}\right)\xi(r) \; ,
\label{eq:xi0} 
\end{equation}
\begin{equation} 
\xi_2 = \left(\frac{4\beta}{3} +
\frac{4\beta^2}{7}\right)[\xi(r)-\overline{\xi}(r)] \; , 
\end{equation}
\begin{equation} 
\xi_4 = \frac{8\beta^2}{35}\left[\xi(r) + \frac{5}{2}\overline{\xi}(r) \; ,
  -\frac{7}{2}\overline{\overline{\xi}}(r)\right],
\end{equation}
where $\xi(r)$ is the real-space {\em undistorted} correlation
function, whereas the {\em barred} functions are:
\begin{equation} 
\overline{\xi}(r) = \frac{3}{r^3}\int^r_0dr'\xi(r')r'{^2} \; ,
\end{equation}
\begin{equation}
\overline{\overline{\xi}}(r) = \frac{5}{r^5}\int^r_0dr'\xi(r')r'{^4} \; .
\label{eq:xi__} 
\end{equation}
Eq.~(\ref{eq:ximodellin}) is a good description of the RSD only at
very large scales, where non-linear effects can be neglected.

A full empirical model, that can account for both linear and
non-linear dynamics, is the so-called ``dispersion model"
\citep{peacock1996, peebles1980, davis1983} in which the redshift-space
correlation function is expressed as a convolution of the
linearly-distorted 
function with the distribution function of pairwise velocities $f(v)$:
\begin{equation} 
 \xi(s_\perp, s_\parallel) = \int^{\infty}_{-\infty}dv f(v)\xi\left(s_\perp, s_\parallel -
 \frac{v(1+z)}{H(z)}\right)_{\rm lin} \; ,
\label{eq:ximodel}
\end{equation}
where the pairwise velocity $v$ is expressed in physical coordinates.

In this paper, we test two different forms for $f(v)$, namely
\begin{equation}
f_{\rm exp}(v)=\frac{1}{\sigma_{12}\sqrt{2}}
\exp\left(-\frac{\sqrt{2}|v|}{\sigma_{12}}\right) \; ,
\label{eq:fvexp} 
\end{equation}
and 
\begin{equation}
f_{\rm gauss}(v)=\frac{1}{\sigma_{12}\sqrt{\pi}}
\exp\left(-\frac{v^2}{\sigma_{12}^2}\right)
\label{eq:fvgauss} 
\end{equation}
\citep{davis1983, fisher1994, zurek1994}.  The quantity $\sigma_{12}$
is independent of pair separations and is generally interpreted as the
dispersion in the pairwise random peculiar velocities. Here we rather
regard it as a free parameter that quantifies the cumulative effect of
small scale random motions and statistical errors on the measured
redshifts $\delta z$, and focus on the distortion parameter $\beta$.

This simple model for RSD depends on a few quantities: two free
parameters, $\beta$ and $\sigma_{12}$, a reference background
cosmology to convert angle and redshifts into distances and the {\em
  true} two-point correlation function of haloes (galaxies) in
real-space, $\xi(r)$, that can be either derived from theory or
estimated from the galaxy redshift catalogue itself.  A theoretical
expression for the galaxy correlation function can be obtained, for
example, from the observed galaxy luminosity function in the framework
of the Halo Occupation Distribution by adopting a theoretical
prescription for the halo two-point correlation function \citep[see
  e.g.][]{yang2003}.  Alternatively, it is possible to estimate
$\xi(r)$ from the measured \xiiz, with a two-step {\em deprojection}
procedure. First, the observed redshift-space correlation function is
projected along $s_\parallel$:
\begin{equation} 
\Xi(r_\perp) \equiv \Xi(s_\perp) = 2 \int_0^{ s_{\parallel}^{\rm max}}
ds'_\parallel \xi (s_\perp, s'_\parallel) \; .
\label{eq:xiproj} 
\end{equation} 

Then, the real-space correlation function can be estimated from the
Abel integral \citep{davis1983, saunders1992}:
\begin{equation} 
\xi(r) = -\frac{1}{r_\parallel}\int^{\infty}_r dr'_\perp
\frac{d\Xi(r'_\perp)/dr_\perp}{\sqrt{r_\perp'^2 - r^2}} \; .
\label{eq:xideproj} 
\end{equation} 
In this paper we adopt this second approach and assess the impact of
deprojection by comparing the results with the ideal case in which we
use $\xi(r)$ measured directly from the mock catalogues.

To evaluate the correlation function in the simulation we use the
\citet{landy1993} estimator:
\begin{equation}
\xi(r)=\frac{HH(r)-2HR(r)+RR(r)}{RR(r)} \; ,
\label{eqn:landy}
\end{equation}
where $HH(r)$, $HR(r)$ and $RR(r)$ are the fraction of halo--halo,
halo--random and random--random pairs, with spatial separation $r$, in
the range $[r-\delta r/2, r+\delta r/2]$ and $\delta r$ is the bin
size.  Since we are interested in estimating $\beta$ at intermediate
scales, we evaluate the correlation function in bins of size $\delta
r=1$ \Mpch out to 50 \Mpch, both in the parallel and perpendicular
directions.  We have checked that pushing our analysis out to
$r\sim100$ \Mpch does not change significantly the results.  Finally,
since to perform the deprojection procedure one needs to specify the
behaviour of $\xi(r)$ at small scales, we have linearly extrapolated
\xii in the range $r_{\parallel}^{\rm min}<1$ \Mpch.  We have verified
that the results do not depend on the extrapolation scheme adopted.


\subsection{Modelling Geometric Distortions}
\label{subsec:GD}

To convert observed redshifts and angular separations into relative
distances one has to assume a background cosmological model that does
not, in general, coincide with the true one. This mismatch induces
asymmetries or anisotropies that, if spotted, can be used to constrain
the background cosmological model itself. This cosmological test,
commonly known as AP test, can be performed using the observed
two-point correlation function.

\begin{figure*}
\includegraphics[width=\textwidth]{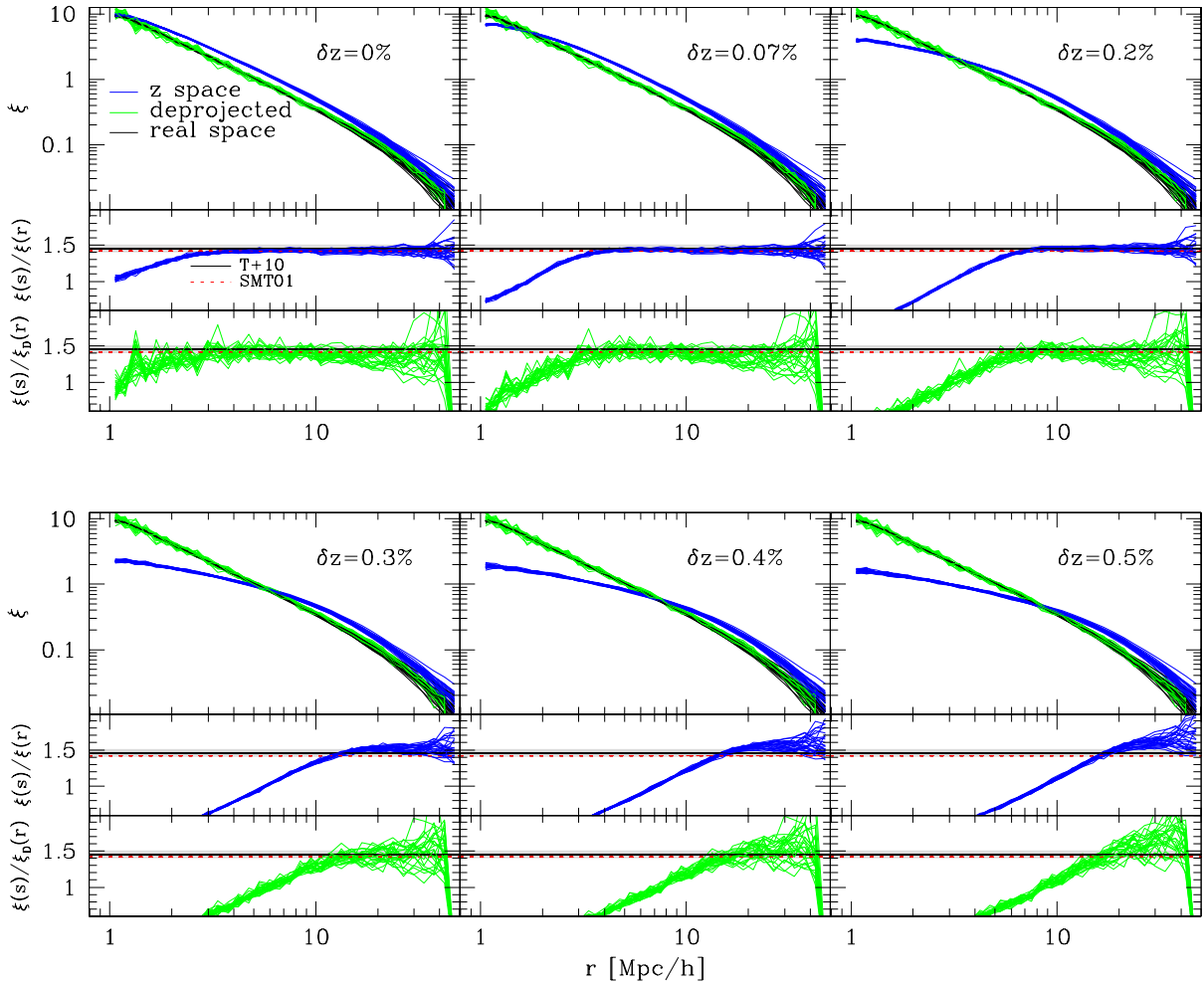}
\caption{The impact of redshift errors on the measured real- and
  redshift-space (angle-averaged) correlation functions.  The upper
  sections of each panel show the real-space {\em true} correlation
  functions $\xi(r)$ (black curves), the redshift-space correlation
  functions $\xi(s)$ (blue curves) and the real-space correlation
  functions obtained from deprojection, $\xi_D(r)$ (green
  curves). Each line shows the correlation measured in one of the 27
  independent sub-boxes (see Section~\ref{sec:tools}). The central and
  middle sections of each panel show the ratios $\xi(s)/ \xi(r)$ and
  $\xi(s)/ \xi_D(r)$, respectively.  Linear theory predictions of
  \citet{tinker2010} and \citet{sheth_mo_tormen2001} are indicated by
  black solid and red dotted lines, respectively.  The grey horizontal
  bands represent $10\%$ theoretical uncertainties.  The panels refer
  to different redshift errors, as indicated by the labels.}
 \label{fig:xi_sigmaV}
\end{figure*}

In absence of peculiar velocities, isotropy in galaxy clustering
guarantees that iso-correlation contours are circles in the
$(r_\perp,r_\parallel)$ plane if pair separations are estimated
assuming the correct geometry.  In this sense, \xii can be considered
a {\em standard circle} in much the same way as BAO and Supernovae are
considered standard rulers and standard candles, respectively. The
choice of an incorrect cosmology will distort these circles in a way
that we can accurately predict.  Indeed, the relation between
  comoving separations in two different geometries reads:
\begin{equation}
r_{\perp 1}=\frac{D_{\rm A,1}(z)}{D_{\rm A,2}(z)} r_{\perp 2}; \; \; r_{\parallel
  1}=\frac{H_2(z)}{H_1(z)} r_{\parallel 2} \; ,
\label{eq:rscal} 
\end{equation}
where the subscripts $1$ and $2$ refer to the two cosmological models
and $D_{\rm A}$ is the angular diameter distance:
\begin{equation}
D_{\rm A}(z) = \frac{c}{H_0(1+z)\sqrt{-\Omega_{\rm
      k}}}\sin\left(\sqrt{-\Omega_{\rm k}}D_{\rm C}(z)\right)  \; .
\label{eq:DA} 
\end{equation}

These relations can be adopted to perform the AP test using the
two-point correlation function as follows.  In the quest for $\beta$,
the choice of a background cosmology is made twice: first to estimate
\xiiz from observed redshifts and angular positions and then to model
\xiiz.  We shall call {\em assumed} and {\em test} cosmology the two
cosmological models assumed to measure and model \xiiz,
respectively. The test is performed by computing $\xi_{\rm
  a}(s_\perp^a,s_\parallel^a)$ for a given {\em assumed} cosmology and
comparing it with the model $\xi_{\rm t}(s_\perp^t,s_\parallel^t)$
estimated for a {\em test} cosmology and rescaled to the assumed one
(Eq.~(\ref{eq:rscal})):
\begin{equation}
\xi_{\rm a}(s_\perp,s_\parallel)=\xi_{\rm t}\left(\frac{D_{\rm
    A,t}}{D_{\rm A,a}}s_\perp,\frac{H_{\rm a}}{H_{\rm
    t}}s_\parallel\right) \; .
\label{eq:xiAP} 
\end{equation}
The correct values of $\beta$, $D_{\rm A}$ and $H$ are found when
$\xi_{\rm t}(s_\perp^t,s_\parallel^t)=\xi_{\rm
  a}(s_\perp^a,s_\parallel^a)$, within the errors.  Unfortunately, the
small amplitude of GD with respect to RSD and inaccuracies in
modelling the latter make results obtained through this procedure not
very robust. Moreover, this method depends explicitly on the bias
model assumed to derive the DM correlation function from the observed
galaxy positions. In this work, we have developed an alternative
procedure, described in Section~\ref{sec:AP}, that exploits both RSD
and GD directly, and does not require modelling the shape of the DM
correlation function and bias.

\begin{figure*}
\includegraphics[width=\textwidth]{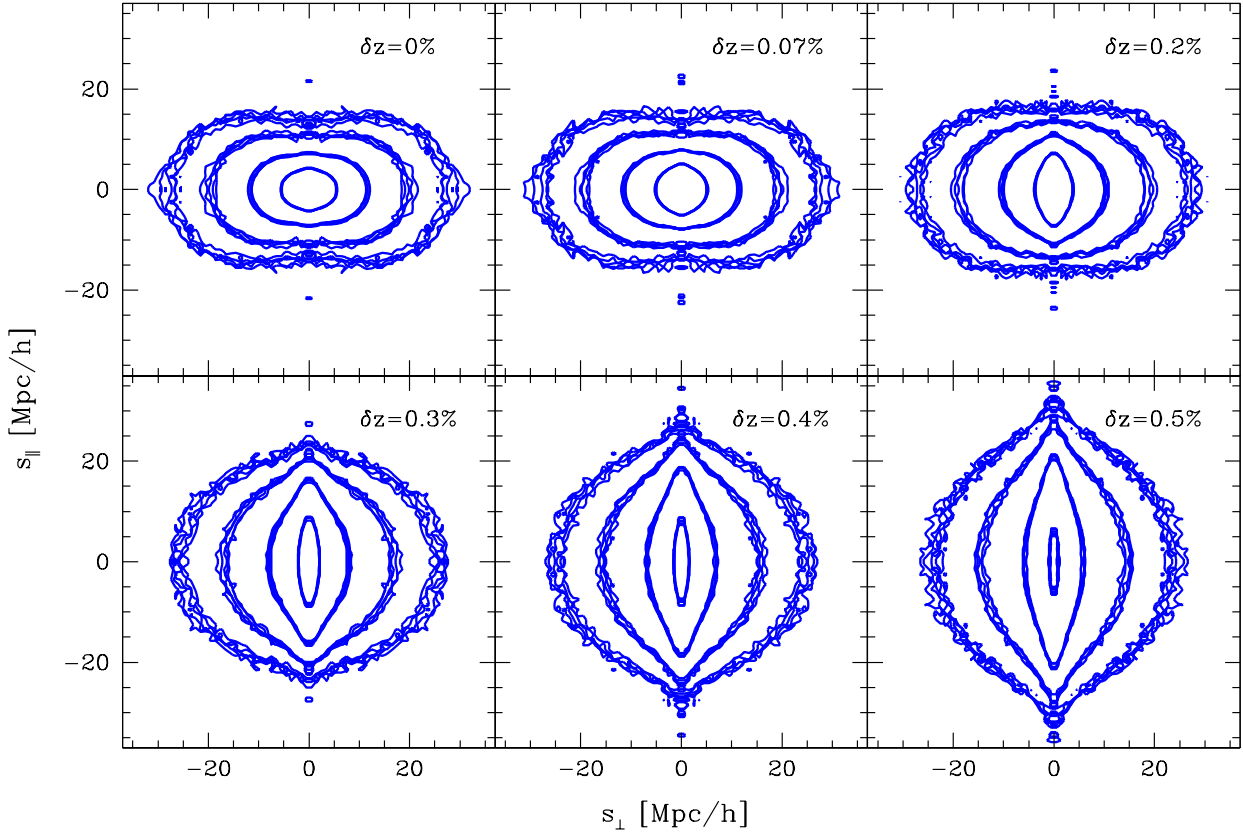}
\caption{Iso-correlation contours of \xiiz measured in the 27 mock
  catalogues. Contours are drawn in correspondence of the correlation
  levels \xiiz$=\{0.1,\,0.2,\,0.5,\,1.5\}$. Different panels refer to
  different amplitudes of the redshift errors, as indicated in the
  labels.}
 \label{fig:iso_sigmaV}
\end{figure*}

\begin{figure*}
\includegraphics[width=0.49\textwidth]{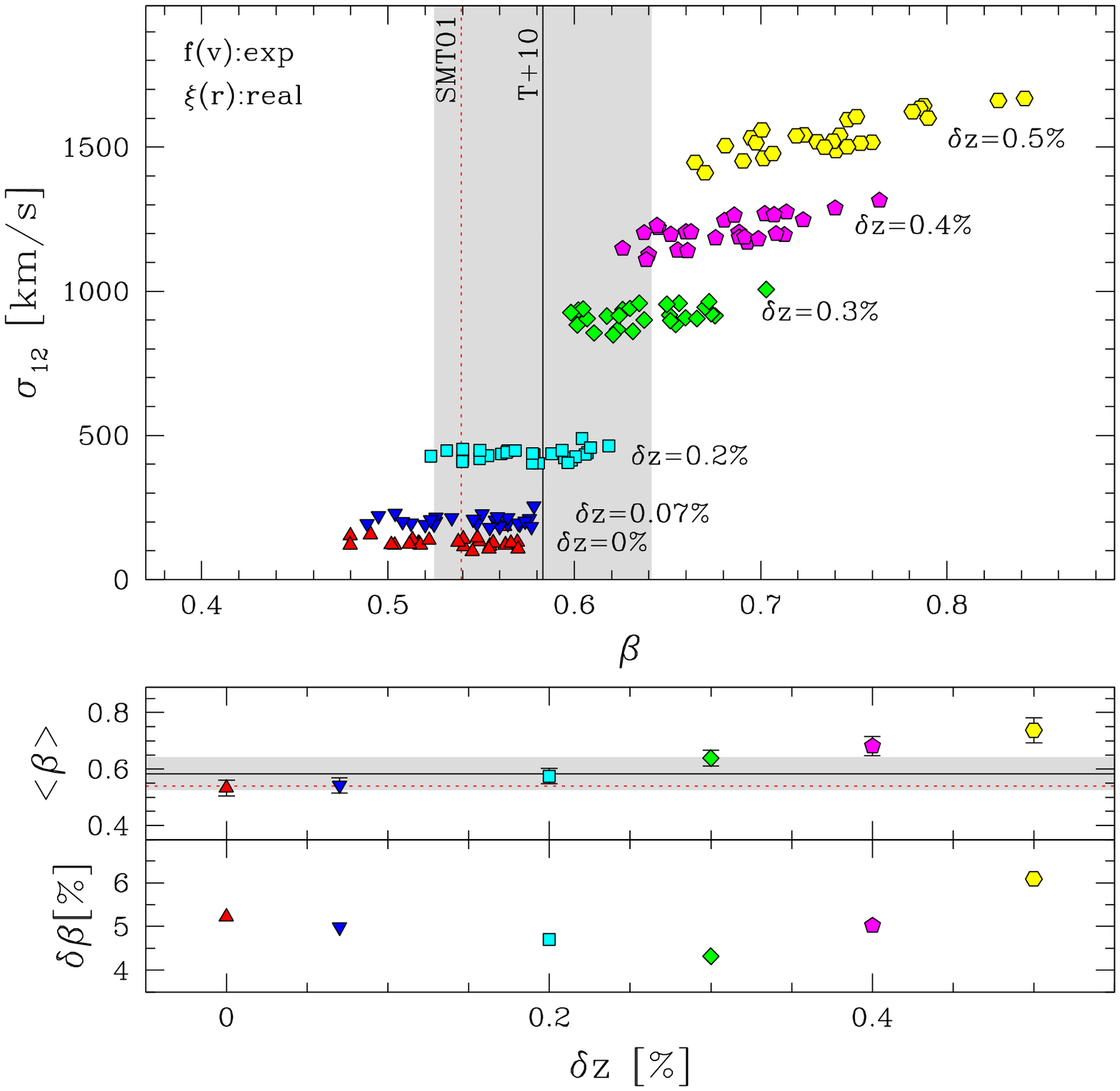}
\includegraphics[width=0.49\textwidth]{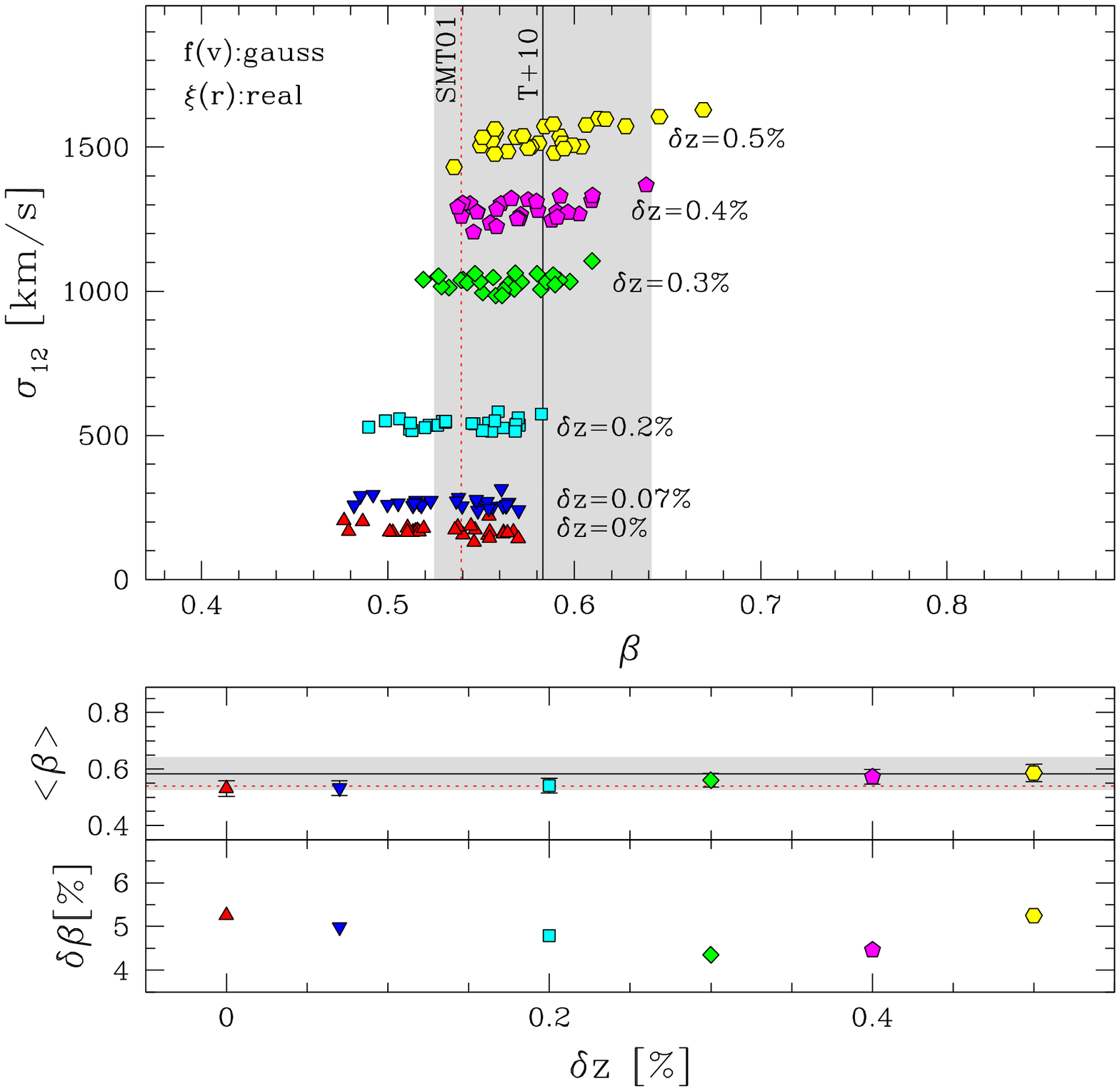}
\includegraphics[width=0.49\textwidth]{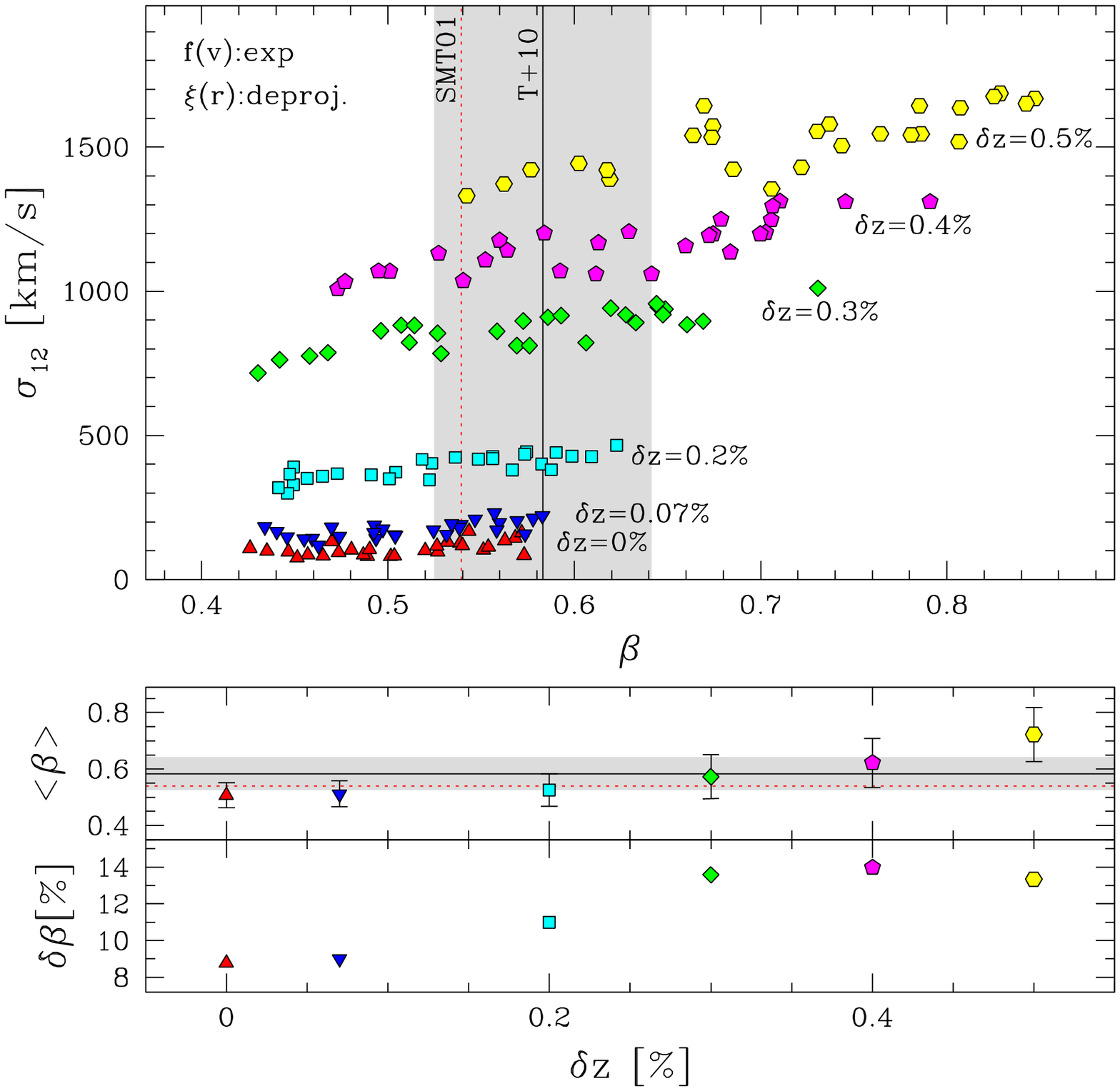}
\includegraphics[width=0.49\textwidth]{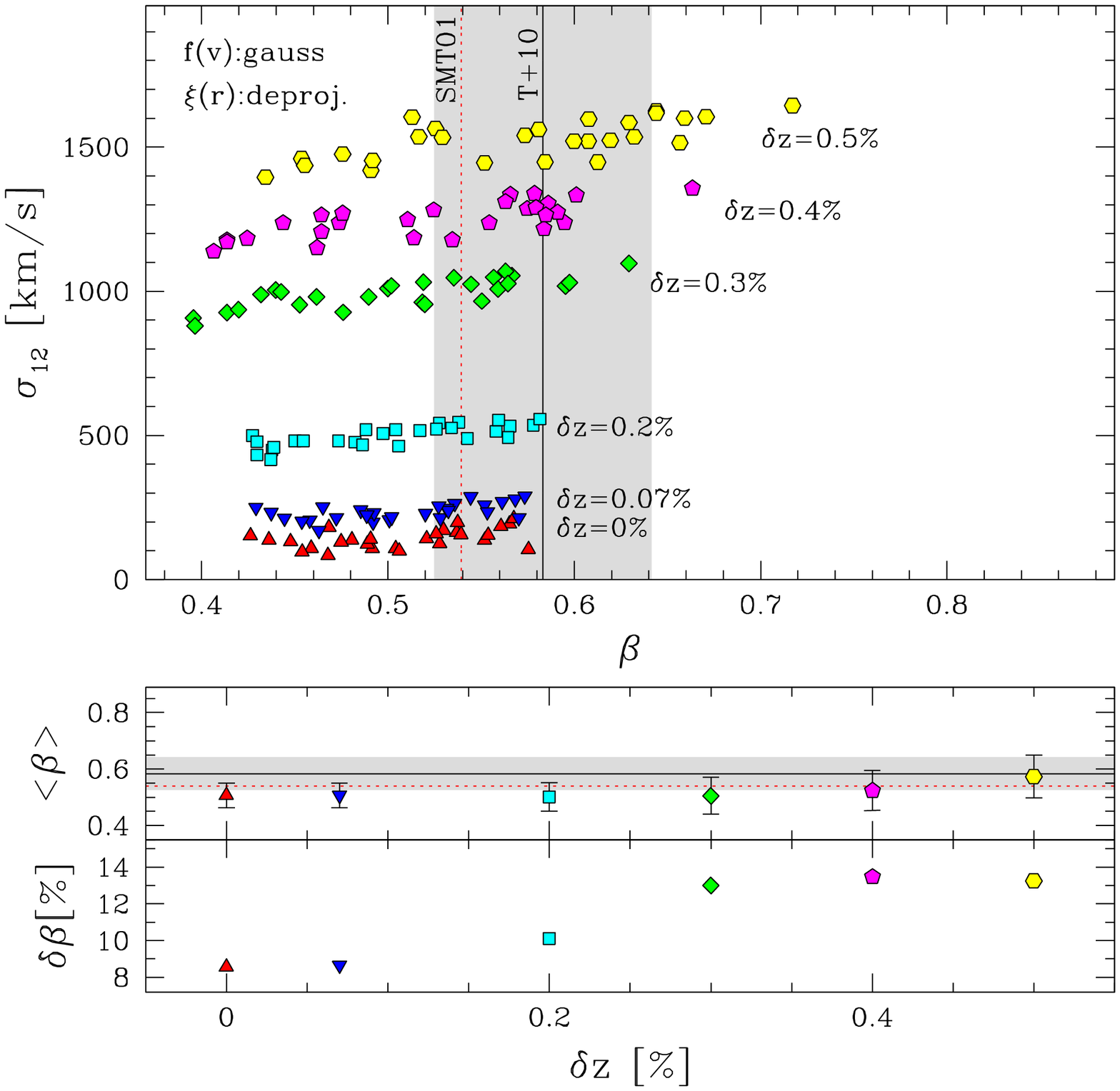}
\caption{{\em Large upper panels}: best-fit parameters $\beta$
    and $\sigma_{12}$, as a function of the redshift error $\delta z$.
    {\em Small bottom panels}: systematic and statistical errors, as a
    function of $\delta z$. Specifically, the upper half windows show
    the best-fit values of $\beta$, averaged over the 27 mocks, and
    their scatter, $\sigma(\beta)$, represented by the error bars.
    The lower half windows show the random component of the relative
    error, $\sigma(\beta)/\beta \, [\%]$.  Black solid and red dotted
  lines represent theoretical expectations from \citet{tinker2010} and
  \citet{sheth_mo_tormen2001}, respectively. The grey bands represent
  a theoretical uncertainty of $10\%$.  The assumption on the adopted
  shape of $f(v)$ and on $\xi(r)$ are labeled in the top-left part of
  the large upper panels. Top-left part of the figure: $f(v)$
  exponential form and {\em true} $\xi(r)$. Top-right: $f(v)$ Gaussian
  form and {\em true} $\xi(r)$.  Bottom-left part of the figure:
  $f(v)$ exponential form and {\em deprojected} $\xi(r)$.
  Bottom-right: $f(v)$ Gaussian form and {\em deprojected} $\xi(r)$.}
\label{fig:beta_sigma12_sigmaV}
\end{figure*}


\section {Measuring $\beta$ from RSD}
\label{sec:rsd}

\subsection{The impact of redshift errors and deprojection}

Random errors in the measured spectroscopic redshifts contaminate the
clustering signal at all scales in a way similar to that of random
peculiar motions at small scales. So far they have not been considered
in the error budget, as local surveys typically had fairly precise
redshift measurements errors ($<100$ km s$^{-1}$).  A careful
assessment of their impact is now in order, especially in view of
future surveys using different observing techniques. For example,
Euclid will be measuring redshifts from single emission lines in
slitless spectroscopic observations, with a requirement on the errors
of $\sigma_z<0.001(1+z)$, which corresponds to $600$ km s$^{-1}$ at
$z=1$ \citep{laureijs2011}. 

To assess the impact of redshift errors on the estimates of $\beta$,
we have perturbed the redshifts of the mock galaxies
(Eq.~(\ref{eq:redshift})) by adding a Gaussian noise of amplitude
$\sigma_v =\{0,\,200,\,500,\,1000,\,1250,\,1500\}$ \kms. These values
cover a range extending out to errors that get close to (yet not as
large as) those from photometric estimates (which in the best cases
have $\sigma_v\sim9000(1+z)$ \kms).  In what follows we express these
errors as per cent uncertainties $\delta
z[\%]=\{0,\,0.07,\,0.2,\,0.3,\,0.4,\,0.5\}$
(Eq.~(\ref{eq:redshift})). 

To focus on the impact of redshift errors we do not consider GD,
i.e. we assume the correct background cosmology.  Moreover, we
  restrict our analysis to the case of Gaussian redshift errors. This
  is a quite accurate assumption for modelling the redshift errors of
  typical spectroscopic galaxy surveys \citep[see e.g.][]{lilly2009}.
  Furthermore, even in photometric galaxy surveys, it has been
  demonstrated that the adoption of a Gaussian distribution to
  simulate the impact of redshift errors represents a reasonable
  approximation \citep[see e.g.][]{cunha2009, saglia2012}. We do not
  consider the impact of the so-called catastrophic errors, the ones
  caused by the misidentification of one or more spectral features,
  for the following reasons. In the assumption that the catastrophic
  outliers represent a perfect isotropic population (i.e. if such
  errors have a flat distribution), their effect is to reduce the
  amplitude of the correlation function at all scales. So they do not
  induce additional distortions in galaxy clustering and do not bias
  the estimate of $\beta$. On the other hand, clustering distortions
  might be generated by systematic misidentification of spectral
  features. Their impact on $\beta$ can only be estimated with mock
  galaxy catalogues, taking into account all observational effects
  that may vary case by case.

\subsubsection{Real- and redshift-space correlation functions}
In Fig.~\ref{fig:xi_sigmaV} we show the redshift-space two-point
correlation functions, $\xi(s)$, measured in the 27 mocks (blue curves
in the upper part of the six panels) and compare them to the
real-space ones, $\xi(r)$ (black curves).  Redshift errors suppress
the clustering amplitude on progressively large scales, as expected.
A first quantitative assessment of this effect can be obtained from
the ratio $\xi(s)/\xi(r)$ which, in the linear limit, is simply
related to $\beta$:
\begin{equation} 
  \frac{\xi(s)}{\xi(r)} = 1 + \frac{2\beta}{3} + \frac{\beta^2}{5} \; .
\label{eq:xiratio} 
\end{equation}
We plot this ratio in the middle part of each panel in
Fig.~\ref{fig:xi_sigmaV} (blue curves).  When $\delta z=0$, this ratio
is constant for $r\gtrsim3$ \Mpch.  On these scales, the value of
$\beta$ obtained from Eq.~(\ref{eq:xiratio}) is consistent with
theoretical expectations of \citet{sheth_mo_tormen2001} and
\citet{tinker2010}, represented by the red dotted and black solid
horizontal lines, respectively. The horizontal grey band is plotted
for reference and represents a theoretical uncertainty of $10\%$.
Increasing redshift errors to $\delta z\sim0.2\%$ has the effect of
suppressing the clustering amplitude on ever larger scales and reduces
the range useful to measure $\beta$ but does not bias its
estimate. For $\delta z\gtrsim0.3\%$ the ratio $\xi(s)/\xi(r)$ is
biased high in the ever shrinking range of scales in which this ratio
is constant, hence inducing a systematic error on $\beta$ obtained
from Eq.~(\ref{eq:xiratio}).

\begin{figure}
\includegraphics[width=0.49\textwidth]{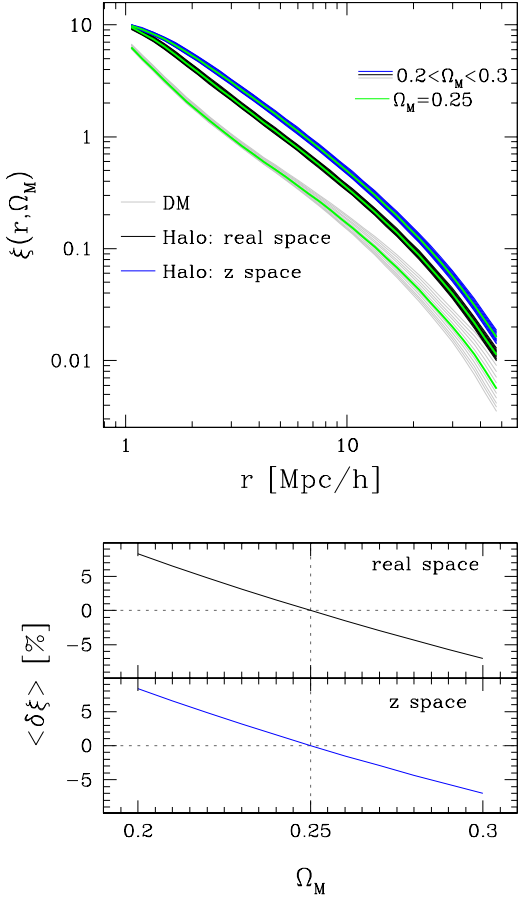}
\caption{Upper panel: effect of GD on the halo correlation function in
  real- (middle set of black curves) and redshift-space (upper
  set of blue curves).  The curves in each set have been obtained
  varying \Om in the range $0.2\leq \Omega_{\rm M}\leq0.3$ (from top
  to bottom).  The lower set of grey curves shows the DM correlation
  function computed with {\small CAMB} \citep{lewis2002}.  The central
  green lines refer to the choice of the correct value $\Omega_{\rm
    M}=0.25$.  Lower panel: mean fractional error on the estimated
  $\xi$ as a function of \Om in real- and redshift-space.}
 \label{fig:xi_OmegaM}
\end{figure}

As a further step towards a realistic estimate of $\beta$, we also
assess the impact of the deprojection procedure described in
Section~\ref{sec:rsd}. The ragged green curves in the upper panels of
Fig.~\ref{fig:xi_sigmaV} show the real-space two-point correlation
function obtained from the deprojection procedure, $\xi_{\rm D}$.  The
corresponding ratio $\xi(s)/\xi_{\rm D}(r)$ is shown in the bottom
panels.  Interestingly, the deprojected correlation function is in
good agreement with the {\em true} $\xi(r)$ even for large values of
$\delta z$, indicating that the deprojection procedure does not
introduce significant systematic errors.  However, it increases random
errors represented by the scatter among the $\xi(s)/\xi_{\rm D}(r)$
curves.

\subsubsection{$\beta$ from the full fit of \xiiz} 
A better estimate of $\beta$ can be obtained by comparing the measured
\xiiz with the model described in Section~\ref{sec:rsd}.  The
iso-correlation contours of \xiiz calculated in the 27 mocks are shown
in Fig.~\ref{fig:iso_sigmaV} for different values of $\delta z$,
indicated in the panels.  Contours refer to the iso-correlation levels
\xiiz$=\{0.1,\,0.2,\,0.5,\,1.5\}$.  The effect of increasing redshift
errors can be clearly appreciated.  The case of no errors ($\delta
z=0$) is characterized by the expected squashing of the contours at
large separations induced by coherent motions whereas on small scales
the fingers-of-God elongation is hardly visible. As already pointed
out, this is due to the lack of substructures in the DM haloes, such
that the velocity field within virialized structures is poorly sampled.
When redshift errors are turned on, fingers-of-God distortions appear
and dominate the distortion pattern out to a scale that increases with
$\delta z$.

Comparing the correlation function ``observed" from the mocks in
Fig.~\ref{fig:iso_sigmaV} with the model presented in
Section~\ref{sec:mod}, constrains the free parameters $\beta$ and
$\sigma_{12}$.  This is done by minimizing the standard $\chi^2$
function:
\begin{equation}
  \chi^2 =
  \sum_{i,j}\frac{\left(y_{ij}-y_{ij}^M\right)^2}{\delta_{ij}^2} \; ,
\label{eq:chi2}
\end{equation}
where $y_{ij}=\xi(s_{\perp,i},s_{\parallel,j})$ and
$y_{ij}^M=\xi^{M}(s_{\perp,i},s_{\parallel,j};\beta,\sigma_{12})$ are
the measured and model correlation functions, respectively, and
$\delta_{ij}=\delta\xi(s_{\perp,i},s_{\parallel,j})$ is the
statistical Poisson noise estimated following \citet{mo1992}. In each
of the 27 mock catalogues, we fitted over the range
$3<r_\perp,r_\parallel<35$ \Mpch, with linear bins of $1$ \Mpch both
in the parallel and perpendicular directions.  As we have explicitly
verified, the results presented in this paper do not depend on the
particular form of Eq.~(\ref{eq:chi2}) and on the definition of
clustering uncertainties $\delta\xi(s_\perp,s_\parallel)$ \citep[for a
  more detailed discussion see][]{bianchi2012}.

\begin{figure}
\includegraphics[width=0.49\textwidth]{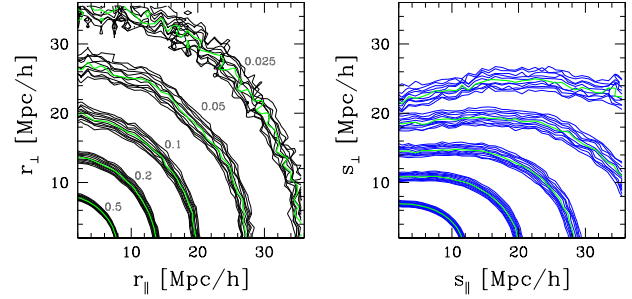}
\caption{Iso-correlation contours for \xii (left panel) and \xiiz
  (right panel). Contours are drawn at the iso-correlation levels
  $\{0.025,\,0.05,\,0.1,\,0.2,\,0.5\}$ indicated by the labels.
  Different contours drawn at the same correlation level refer to the
  different values of \Om considered.  The green contours refer to the
  correct cosmological model, \Om=0.25.}
 \label{fig:iso_OmegaM}
\end{figure}

The results are summarized in Fig.~\ref{fig:beta_sigma12_sigmaV}.  Let
us focus on the upper left part of the figure. The points in the top
panel represent the best-fit values of $\beta$ and $\sigma_{12}$
obtained from each mock catalogue. The different symbols and colours
indicate different redshift errors $\delta z$, as specified in the
labels.  The best-fit $\beta$ values should be compared with
theoretical expectations using the \citet{tinker2010} model (black
solid vertical line), which, as discussed previously, is a very good
description of the intrinsic linear bias of our simulated haloes. The
\citet{sheth_mo_tormen2001} model (red dotted vertical line) is shown
for comparison. The vertical grey band shows the $10\%$ uncertainty
interval.  These results have been obtained by comparing data with a
model in which we have used an exponential form for $f(v)$ and the
{\em true} $\xi(r)$ of the DM haloes in the N-body
simulation. Systematic and statistical errors and their dependence on
$\delta z $ are quantified in the bottom panel. In the upper part we
show the mean value of the best-fit $\beta$ obtained by averaging over
the 27 mock catalogues and its scatter, $\sigma(\beta)$, represented
by the error bars. In the bottom panel we show the random component of
the relative error $\sigma(\beta)/\beta \, [\%]$.

\begin{figure}
\includegraphics[width=0.49\textwidth]{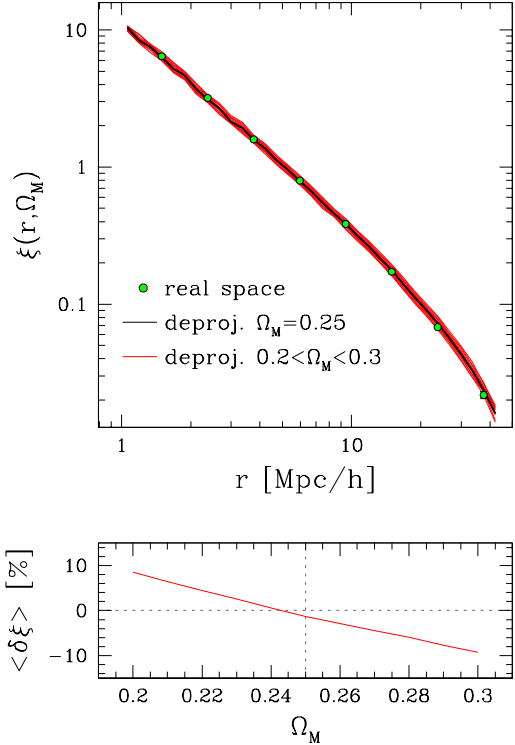}
\caption{Upper panel: {\em true} correlation function in real-space
  (green dots), deprojected correlation function with \Om=0.25 (black
  curve) and deprojected correlation function with other values of \Om
  (red curves).  Bottom panel: mean fractional error $<\delta\xi>$ as
  a function of \Om.  }
 \label{fig:xi_OmegaM_deproj}
\end{figure}

As shown in \citet{bianchi2012}, such systematic error depends on the
minimum mass (i.e. the bias) of the haloes considered and tends to
decrease up to masses of $10^{13} \mbox{M}_\odot$.  In
Fig.~\ref{fig:beta_sigma12_sigmaV} we see, however, that redshift
errors larger than $\delta z\sim0.3\%$ produce an opposite effect,
which cancels and then overcomes the intrinsic negative systematic
bias on $\beta$.  Interestingly, the {\it rms} error remains instead
substantially constant, when the real-space correlation function is
well known (upper panel, $\sim 5\%$ for the volumes considered here).

The upper right part of the same figure shows the results obtained
when we model the velocity distribution function $f(v)$ with a
Gaussian function instead of an exponential one. The main effect is a
very significant reduction of the systematic errors. This is due to
the fact that redshift errors, modelled as Gaussian variables, can be
regarded as a random velocity field with a distribution function that
obeys a Gaussian statistics.  What we learn here is that when redshift
errors dominate over the pairwise velocity dispersions, then $f(v)$ is
best modelled by a Gaussian function with dispersion comparable to
$\delta z$. This is demonstrated by the fact that, in the plot, the
best-fit values of $\sigma_{12}$ are comparable to the amplitude of
the input redshift errors when $\delta z$ is large.

The plots in the bottom part of Fig.~\ref{fig:beta_sigma12_sigmaV} are
analogous to those shown in the upper half except for the fact that,
in this case, we are considering the more realistic scenario in which
$\xi(r)$ is not known {\em a priori} but obtained from deprojection.
Uncertainties in the deprojection procedure increase random errors by
a factor of 2-3, depending on the amplitude of $\delta z$.


\subsection {The impact of geometric distortions}

Before looking in more details into how GD arising from the choice of
a wrong cosmological background can actually be exploited to our
benefit, we would like first to understand how they impact the
measurements of the growth rate from RSD. We first investigate how GD
affect the estimate of the correlation function and galaxy bias. We
then focus on the measurement of $\beta$.  Specifically, we assume a
flat cosmology (so that $\Omega_\Lambda=1-$\Om) and investigate the
effect of choosing an incorrect value of \Om in the range $[0.2,0.3]$,
in steps of $\Delta \Omega_{\rm M}=0.01$.  All the other cosmological
parameters are kept fixed to their true values.  For this set of
experiments we set redshift errors $\delta z=0$.  Since the amplitude
of GD is smaller than that of RSD, to appreciate their impact we need
to minimize sampling errors, i.e. trace velocities and density
fluctuations with as many haloes as possible.  Thus, in the following
we shall use the whole simulation box with all its haloes, instead of
the 27 subsamples.

\begin{figure}
\includegraphics[width=0.49\textwidth]{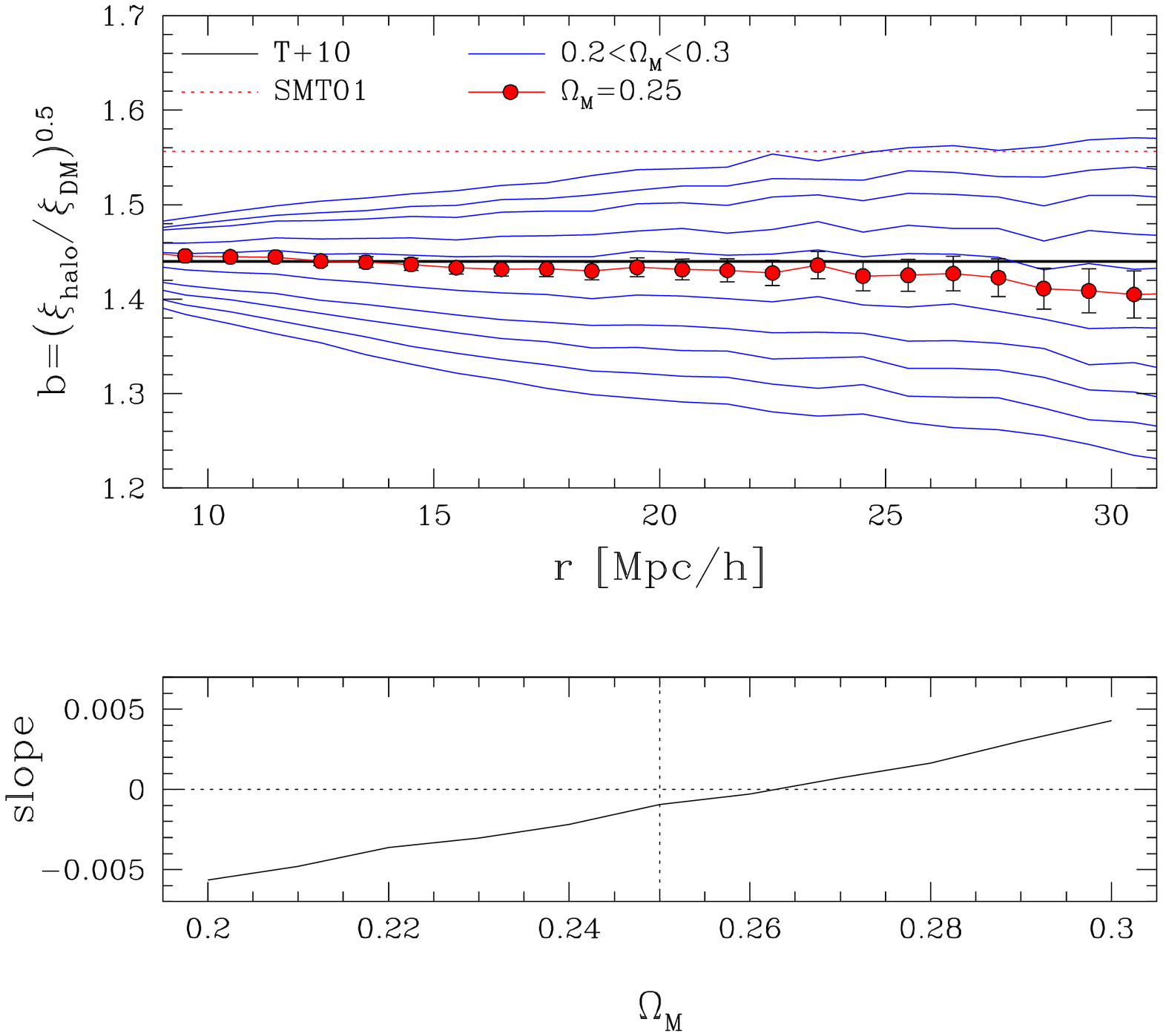}
\caption{Upper panel: bias of the DM haloes in the simulation,
  $b=(\xi_{\rm halo}/\xi_{\rm DM})^{0.5}$, as a function of
  separation, estimated for different values of \Om in the range
  $0.2\leq\Omega_{\rm M}\leq0.3$ (blue curves). Red dots highlight
  the case of a correct $\Omega_{\rm M}=0.25$.  Error bars represent
  statistical uncertainties \citep{mo1992}.  Expected values from the
  models of \citet{tinker2010} and \citet{sheth_mo_tormen2001} are
  represented by the horizontal lines.  Bottom panel: mean slope of
  $b(r)$ in the range $10< r <30$ \Mpch.}
\label{fig:bias:OmegaM}
\end{figure}


\subsubsection{Impact on the measured correlation function}

Fig.~\ref{fig:xi_OmegaM} shows the effect of GD on the measured
two-point halo correlation function in real- (middle set of black
curves) and redshift-space (upper set of blue curves).  The lower set
of grey curves represents the correlation function of the DM obtained
by Fourier transforming the matter power spectrum computed with
{\small CAMB} \citep{lewis2002}, which exploits the HALOFIT routine
\citep{smith2003}.  In each set, the central, green curve refers to
the correct choice of background cosmology, \Om=0.25.  The other
curves refer to \Om values ranging from 0.2 (top) to 0.3 (bottom).
The choice of the incorrect cosmology also distorts the shape of the
computational box. To account for this spurious effect, the random
objects used to compute $\xi(r)$ have been generated within the same,
distorted volume.

\begin{figure}
\includegraphics[width=0.49\textwidth]{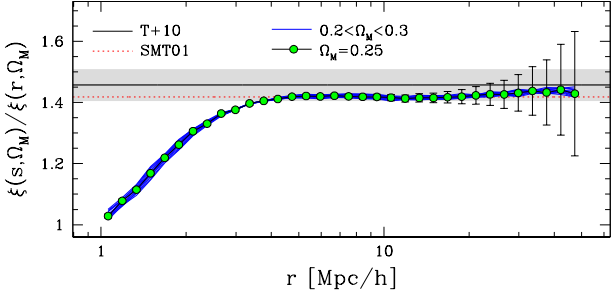}
\caption{Plot of the ratio between the redshift- and real-space
  correlation functions, computed from the simulation box after
  converting redshifts into distances using 10 different cosmologies,
  corresponding to 10
  values of \Om in the range $[0.2,0.3]$ (blue curves). The
  green dots mark the correct value, for \Om=0.25, whereas the horizontal
  lines show theoretical predictions from \citet{tinker2010} (black
  solid) and \citet{sheth_mo_tormen2001} (red dotted). The grey band
  represents a $10\%$ uncertainty around the Tinker et al. value.}
\label{fig:xi_OmegaM_ratio}
\end{figure}

GD enhance/dilute the correlation signal on all scales and thus modify
the amplitude but not the shape of the correlation function. The
effect, quantified by the width of each set of curves, is very
small. It can be better appreciated in the bottom panel in which we
plot the mean fractional residual of $\xi$, $<\delta\xi>
[\%]=\left<(\xi(\Omega_{\rm M})-\xi(\Omega_{\rm
  M}=0.25)/\xi(\Omega_{\rm M}=0.25)\right>$, where the average is over
the interval $1< r <50$ \Mpch.  Since to first-order GD do not modify
the shape of $\xi(r)$, the value of $<\delta\xi> $ quantifies the
amplitude of the spurious boost in the correlation signal induced by
GD.  In correspondence to the values \Om=0.2 and 0.3, already
almost excluded by current observational constraints, the boost is
$\sim 8\%$.

Fig.~\ref{fig:iso_OmegaM} shows the effect of GD on \xii (left panel)
and \xiiz (right panel). Contours are drawn at the correlation values
$\{0.025,\,0.05,\,0.1,\,0.2,\,0.5\}$. The different curves at a given
correlation level refer to different values of \Om. The green contours
refer to the true geometry.

Fig.~\ref{fig:xi_OmegaM_deproj} demonstrates that GD have little
impact on the deprojection procedure.  The green dots show the {\em
  true} correlation function in real-space.  The black curve shows the
deprojected correlation function obtained assuming the correct value
of \Om. The other red curves refer to the other choices of \Om. The
effect is the same as in Fig.~\ref{fig:xi_OmegaM}: an incorrect value
for \Om boosts up or down the correlation amplitude on all scales by a
factor $\lesssim 10\%$ (bottom panel), similarly to when $\xi(r)$ is
measured directly.

\begin{figure}
\includegraphics[width=0.49\textwidth]{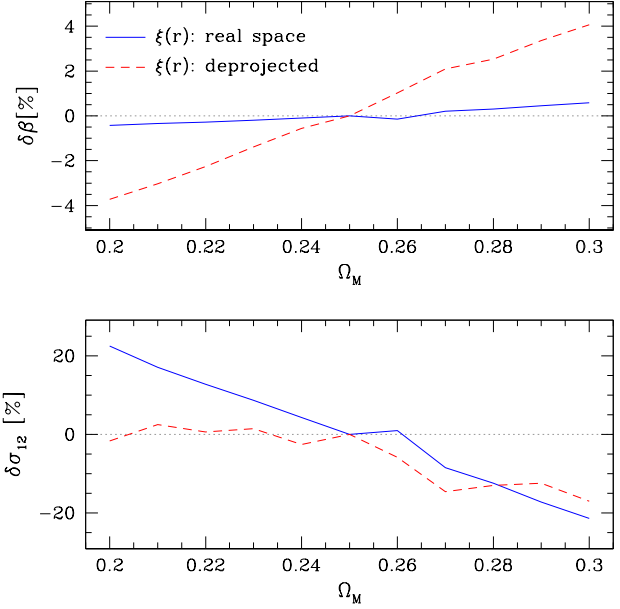}
\caption{Per cent mismatch between the value of $\beta$ (upper panel)
  and $\sigma_{12}$ (lower panel) computed for different values of \Om
  and the one obtained assuming the correct \Om: $\delta X=
  [X(\Omega_{\rm M})-X(\Omega_{\rm M}=0.25)]/X(\Omega_{\rm M}=0.25)$,
  where X can be either $\beta$ or $\sigma_{12}$.  The curves show the
  result obtained assuming the true $\xi(r)$ (solid, blue) or the
  deprojected one (dashed, red).}
\label{fig:beta_sigma12_OmegaM}
\end{figure}

\begin{figure*}
\includegraphics[width=\textwidth]{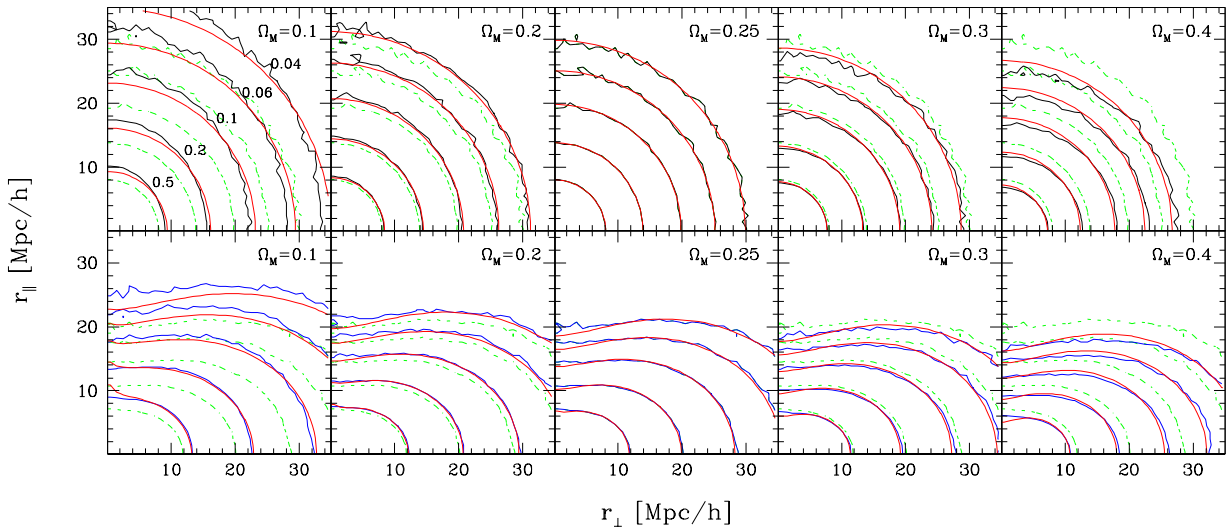}
\caption{Contour plots of the DM halo correlation function, \xii (blue
  curves), in real (top panels) and redshift space (bottom panels),
  measured for $\Omega_{\rm M}=\{0.1,0.2,0.25,0.3,0.4\}$.  The
  iso-correlation contours correspond to
  \xii$=\{0.04,\,0.06,\,0.1,\,0.2,\,0.5\}$.  The red lines in both
  sets of panels correspond to the model \xiiz. In the top panels,
  they simply give the dynamically-undistorted real-space $\xi(r)$
  computed for the given cosmology and replicated over $\pi/2$; in the
  bottom ones, they also include the RSD dispersion model
  (Eq.~(\ref{eq:ximodel})) for the best-fit values of $\beta$ and
  $\sigma_{12}$ derived for that specific \Om. The green dotted curves
  show the {\em geometrically undistorted} \xii measured at the true
  cosmology ($\Omega_{\rm M}=0.25$), for comparison.}
 \label{fig:iso_model_OmegaM} 
\end{figure*}


\subsubsection{Impact on the measured galaxy bias}

An interesting aspect of GD is that the two-point correlation function
estimated assuming an incorrect value $\widetilde{\Omega}_{\rm M}$ is
different from the correct two-point correlation function measured in
a universe with $\widetilde{\Omega}_{\rm M}$.  This fact has some
practical consequences, for example in the measurement of galaxy
bias. Estimates of galaxy bias can be obtained from the ratio of the
galaxy and mass two-point correlation functions.  For example, the
bias of the haloes can be estimated as $b_{\rm halo}(r)=(\xi_{\rm
  halo}(r,\Omega_{\rm M})/\xi_{\rm DM}(r,\Omega_{\rm M}))^{0.5}$,
where $\xi_{\rm halo}(r,\Omega_{\rm M})$ is the real-space halo
correlation function measured assuming some value of \Om and $\xi_{\rm
  DM}(r,\Omega_{\rm M})$ is the mass correlation function in the same
cosmology.  We have computed $b_{\rm halo}(r)$ for the haloes in our
mock catalogues. Results are shown in the upper panel of
Fig.~\ref{fig:bias:OmegaM}, in which we show $b_{\rm halo}(r)$
obtained for different values of \Om in the range $0.2\leq\Omega_{\rm
  M}\leq0.3$ (blue curves, from bottom to top).  Red dots refer to the
correct cosmology and error bars represent 1$\sigma$ statistical
uncertainties computed as in \citet{mo1992}.  Horizontal lines show
the model predictions of \citet{tinker2010} and
\citet{sheth_mo_tormen2001}.

These results show that GD affect both the amplitude of the estimated
bias and its scale dependence.  To estimate the effect, we fit each
curve in the plot with a power law function $b(r)=A+\gamma\cdot r$.
The spurious scale dependence is quantified by the slope $\gamma$ that
we plot in the bottom panel as a function of \Om.  Ideally, it would
seem possible to estimate \Om by requiring that $b(r)$ remains flat on
those scales where it should be constant. However, the smallness of
the effect and the theoretical uncertainties on galaxy bias prevent
this technique to be applied to real data.


\subsubsection{Impact on the measured value of  $\beta$}

To assess the GD impact on $\beta$, we have repeated the same analyses
presented in Section~\ref{sec:rsd}, i.e. we have estimated $\beta$
from the ratio $\xi(s)/\xi(r)$ and by fitting the full \xiiz.  The
blue curves in Fig.~\ref{fig:xi_OmegaM_ratio} show the ratio between
the real- and redshift-space correlation functions for 10 different
values of \Om in the range $0.2\leq\Omega_{\rm M}\leq0.3$ (from bottom
to top).  The green dots refer to the true cosmology case. Error bars
show the statistical errors computed according to the \citet{mo1992}
prescription. Reference values according to \citet{tinker2010} and
\citet{sheth_mo_tormen2001} are shown by the black solid and the red
dotted lines, respectively, with a $10\%$ theoretical uncertainty
indicated by the grey band.  The scatter among the blue curves is
significantly smaller than theoretical uncertainties, thus indicating
that estimates of $\beta$ from RSD in the range
$3<r[h^{-1}\,\mbox{Mpc}]<50$ are robust to the choice of \Om.

The alternative way to estimate $\beta$ from \xiiz confirms this
result.  Fig.~\ref{fig:beta_sigma12_OmegaM} quantifies the amplitude
of the effect.  In the upper panel we show the per cent difference
between $\beta$ computed for a given \Om, indicated on the X-axis, and
the one obtained assuming the correct model \Om=0.25. The solid blue
curve refers to the case in which we model \xiiz using the true
$\xi(r)$, whereas the red dashed curve shows the case in which $\xi(r)$
has been obtained by deprojecting \xiiz.  In both cases, the impact of
GD is rather small, especially if compared to that of redshift
errors. The corresponding error induced on $\beta$ is less than $1\%$
over the whole range of \Om analysed, when the true $\xi(r)$ is used
in the model. Using the deprojected $\xi(r)$, the maximum error on
$\beta$ rises to $\sim 4\%$.  Finally, the lower panel shows the
impact of GD on the other parameter of the fit, the pairwise
dispersion $\sigma_{12}$. The error on this parameter turns out to be
larger than the one on $\beta$, rising to $\sim 20\%$ for the extreme
values of \Om.


\section {Disentangling dynamic and geometric distortions}
\label{sec:AP}

In this section, we investigate the possibility to perform the AP test
using \xiiz. The goal is to constrain both $\beta$ and the
cosmological parameters that enter Eq.~(\ref{eq:Hubble}) exploiting
anisotropies in galaxy clustering induced by RSD and GD.

\subsection {The method}
\label{subsec:AP0}

As we have anticipated in Section~\ref{subsec:GD}, a common approach
is to use Eq.~(\ref{eq:xiAP}) to model the two-point correlation
function in a series of {\em test} cosmological models and compare it
with the measured one \citep{ballinger1996}.  Here we adopt an
alternative procedure, which is a generalization of the iterative
method introduced in \citet{guzzo2008} and that is found to be robust.
The method consists in repeating the measurement of the correlation
function in different test cosmologies, and then modelling {\em only}
its RSD. Our working hypothesis is that, by construction, the
agreement between model and data will be maximum when the test
cosmology coincides with the {\em true} cosmology of the Universe,
i.e. without GD.

The steps of the procedure can be summarized as follows.
\begin{enumerate}
\item Choose a cosmological model to convert angular positions and
  redshifts into comoving coordinates.  
\item Measure \xiiz.
\item Estimate the real-space correlation function, $\xi(r)$, required
  to model dynamic distortions (e.g. through the deprojection
  technique).
\item Model {\em only} dynamic distortions (e.g. through
  Eq.~\ref{eq:ximodel}), and derive the best-fit values of $\beta$ and
  $\sigma_{12}$ that minimize the $\chi^2$ function given by
  eq.~\ref{eq:chi2}.  
\item Save this specific minimum value of the $\chi^2$, that we shall
  call $F(\{D_{\rm A},H\}_i)$.
\item Go back to point (i) using a different test cosmology and
  estimate a new value for $F$.
\end{enumerate}
Once the whole set of $\{D_{\rm A},H\}_i$ has been explored, the
``best of the best'' set of parameter values
$(\beta,\sigma_{12},D_{\rm A},H)$ will be then identified by the
minimum value of $F(\{D_{\rm A},H\}_i)$.  The main differences between
this procedure and the usual one are that i) the observed and model
correlation functions assume the same {\em test} cosmological model,
and ii) once $D_{\rm A}$ and $H$ are fixed, one only needs to model
RSD. In the case of a flat $\Lambda $CDM background, the success
of this strategy is guaranteed by the small covariance between \Om and
$\beta$ \citep[see e.g.][]{ross2007}. As a consequence, one can obtain
an unbiased estimate of \Om even for an incorrect choice of $\beta$
and $\sigma_{12}$.

One advantage of our procedure is that it does not require the
modelling of the galaxy bias. Since tha galaxy correlation function
can be obtained directly from the data through the deprojection
technique, it is not necessary to model the shape of the DM
correlation function (at point (iii)). The only assumption of the
method is the intrinsic isotropy of the clustering.

One disadvantage is the computational cost, since one has to estimate
\xiiz for each cosmological model to test.  However, the use of
optimized linked-list-, Tree- and FFT-based algorithms allows \xiiz to
be computed sufficiently fast, as to efficiently explore the parameter
space without resorting to supercomputing facilities. Alternatively,
instead of directly measuring the correlation function at different
test cosmologies, it is actually sufficient to measure \xiiz in a
fiducial cosmology and rescale the result to a test cosmology, using
Eq.~(\ref{eq:xiAP}). A second disadvantage is related to the estimate
of the errors.  The best-fit parameters are found by minimizing a
function $F$ that does not obey a $\chi^2$ statistics.  The reason is
that the data themselves, which in this case coincide with the
measured \xiiz, depend on $D_{\rm A}$ and $H$. Therefore, since the
values of $F$ evaluated at different $D_{\rm A}$ and $H$ do not refer
to the same dataset, the function $F$ does not follow a $\chi^2$
statistics. As a consequence, errors on $D_{\rm A}$ and $H$ have to be
evaluated in a different way, as we shall see below.

\begin{figure}
\includegraphics[width=0.49\textwidth]{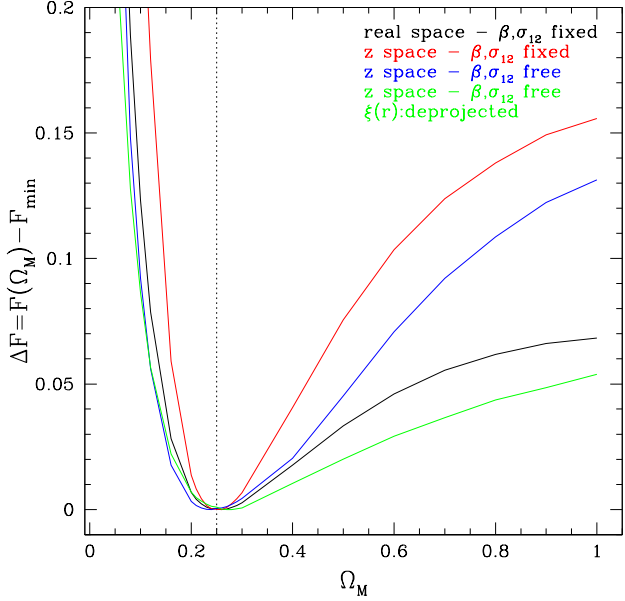}
\caption{The ``pseudo-likelihood'' function $\Delta F$ as a function
  of \Om, obtained with the method described in
  Section~\ref{subsec:AP1}. {\em i}) In real-space fixing
  $\beta=\sigma_{12}=0$ (black line). {\em ii}) In redshift-space, but
  fixing $\beta$ and $\sigma_{12}$ to their best-fit values as derived
  with the correct cosmology (red line). {\em iii}) Same, but leaving
  also $\beta$ and $\sigma_{12}$ as free parameters, either assuming
  perfect knowledge of the true real-space $\xi(r)$ (blue line) or
  {\em iv}) recovering it through deprojection (green line).}
 \label{fig:chi2}
\end{figure}

\subsection {Joint constraints on \Om and $\beta$}
\label{subsec:AP1}

We start our analysis considering the case of a flat $\Lambda $CDM
model in which \Om, the mass density parameter, fully characterizes
the expansion history and geometry of the Universe.
Fig.~\ref{fig:iso_model_OmegaM} illustrates the result of applying
this procedure to the full catalogue of DM haloes.  The different
panels show the iso-correlation contours of the two-point correlation
function measured in real- (black curves in the top panels) and
redshift-space (blue curves in the bottom panels).  Contours are drawn
at the correlation levels $\{0.04,\,0.06,\,0.1,\,0.2,\,0.5\}$.
Different panels refer to the different values of \Om used to compute
distances and estimate the correlation function, as indicated by the
labels.  The green dotted curves are drawn for reference and show the
predictions for the true cosmological model \Om=0.25.  As such, in the
central panel they coincide with the black and blue curves.  The red
curves show the corresponding model for the two-point correlation
function obtained using the best-fit values of $\beta$ and
$\sigma_{12}$ estimated at each value of \Om.

In real-space (top panels) the model for \xii is simply a replica over
all angles of the real-space correlation function $\xi(r)$ (i.e. no
RSD are present, corresponding to setting $\beta=\sigma_{12}=0$ in the
dispersion model). This is shown here to evidence the interplay of the
two effects. It could be seen as an idealized case in which we are
able to perfectly correct for RSD, or can hypothetically reconstruct
the real-space galaxy distribution.  The iso-correlation contours are
thus circles in the $(r_\perp,r_\parallel) $ plane, when the correct
cosmology is used.  The effect of GD when varying the cosmology is
then quantified by the mismatch between the green and the black
contours. As evident, the best-fit value for \Om can be found by
minimizing the difference between the red and the black curves, which
is in practice the AP test.  The best agreement is found for \Om=0.25,
as expected, showing that this procedure is unbiased.

Similar considerations can still be applied to the redshift-space case
(bottom panels).  For a given \Om, the amplitude of the mismatch is
similar to that found in real-space. This fact validates the
hypothesis that GD and RSD are substantially independent.  The
difference between red and blue curves is still minimized for the
correct reference value, \Om=0.25, showing that the result is unbiased
also in redshift-space.

\begin{figure*}
\includegraphics[width=\textwidth]{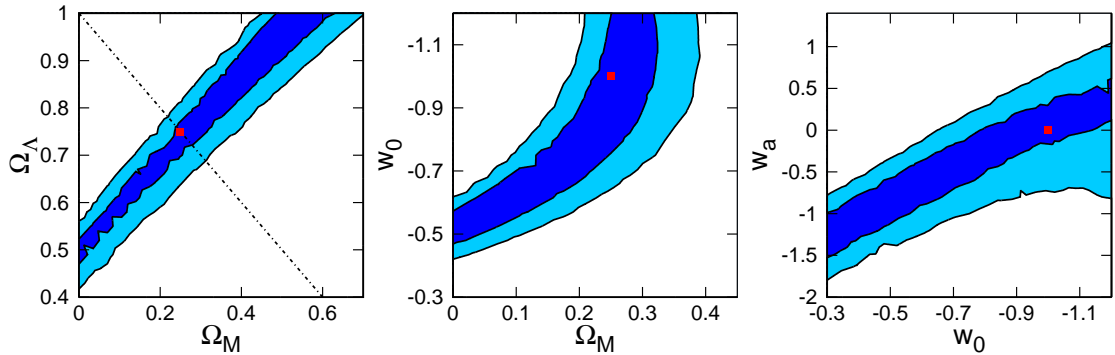}
\caption{The 68 and 95 per cent pseudo-likelihood probability contours
  in the $\Omega_{\rm M}-\Omega_\Lambda$ (left panel), $\Omega_{\rm
    M}-w_0$ (central panel) and $w_0-w_a$ (right panel) planes. In
  each case, the other two parameters not shown are fixed to their
  true values. These constraints have been obtained in redshift-space,
  using the {\em true} $\xi(r)$ and fixing $\beta$ and $\sigma_{12}$
  to their best-fit values as derived with the correct cosmology. The
  red squares mark the correct cosmological parameters of the
  simulation.  The dotted black line in the left panel shows the flat
  background case: $\Omega_\Lambda=1-\Omega_{\rm M}$.}
 \label{fig:APpar}
\end{figure*}

Let us then quantify the ability of the proposed technique to jointly
estimate \Om and $\beta$.  As we described, the best-fit values for
$(\beta,\sigma_{12},\Omega_{\rm M})$ are found at the minimum of the
pseudo--$\chi^2$ function, $F$.  Note that, in this procedure, both
the model and the measured \xiiz depend on \Om.  The same happens with
the errors, since the number of pairs in each bin is modified by the
presence of GD. However, we have verified that this effect is small
and can be ignored. More specifically, the shape of the function $F$
and the position of its minimum are very insensitive to
$\delta\xi(\Omega_{\rm M})$. In Fig.~\ref{fig:chi2} we plot $\Delta
F=F(\Omega_{\rm M})-F_{\rm min}$, where $F_{\rm min}$ is the minimum
value of $F$ found during the exploration of the cosmological
parameter grid (\Om$_i$ in our case). As in Section~\ref{sec:rsd}, we
fit over the range $3<r_\perp,r_\parallel<35$ \Mpch, with linear bins
of $1$ \Mpch both in the parallel and perpendicular directions.

To evidence how the technique operates in detail, and understand its
possible limitations, we proceed in increasing steps, as we did in
Fig.~\ref{fig:iso_model_OmegaM}.  We therefore first test the validity
of the best-fitting procedure in real-space, an ideal case that would
correspond to a perfect subtraction (or absence) of RSD.  In this
case, the value of $F$ quantifies the mismatch between the black and
the red contours shown in the top panels of
Fig.~\ref{fig:iso_model_OmegaM}.  The corresponding function $\Delta
F$ is represented by the black curve in Fig.~\ref{fig:chi2}.  The
minimum of the curve is found for \Om= 0.25, again showing that the
fitting procedure gives unbiased results.

The red curve refers to redshift-space. It shows the values of $\Delta
F$ obtained as a function of \Om$_i$, after fixing $\beta$ and
$\sigma_{12}$ to the best-fit values computed at $\Omega_{\rm
  M}=0.25$.  In practice, this is again an idealized case in which the
correct distortion pattern is known a priori and used as a reference
against the one observed when assuming different cosmologies.  The RSD
model is, in other words, inaccurate for all choices of \Om but for
$\Omega_{\rm M}=0.25$.  Also in this case, the minimum of $\Delta F$
is found for $\Omega_{\rm M}=0.25$, thus indicating that the switch-on
of RSD does not bias our estimate. Interestingly, in this case the
minimum is sharper than in the real-space case (black curve); this can
be explained as due to the stronger constraints posed by the RSD
pattern in the observed \xii, which reduces simmetry with respect to
the simple real-space case.  In other words, this is telling us that,
if we were able to know redshift distortions perfectly, e.g. from an
independent measurement, then the constraints on the background
cosmological parameters from an AP test would be more precise than
those expected from the standard real-space geometric test.  This is
shown in this simplified case by the fact that the red curve yields a
smaller uncertainty on \Om than the black one.

In the general case, however, we do not know a priori the amount of
redshift distortions and we would rather like to estimate also $\beta$
(and $\sigma_{12}$), together with \Om.  The resulting constraint is
shown by the blue and green curves.  For the blue curve, we have
assumed perfect knowledge of the real-space correlation function
$\xi(r)$ (i.e. we have measured it directly from real-space positions
in the simulation). We have already seen how crucial this is, as an
ingredient in the RSD model.  Also in this case, the minimum is found
at the expected value $\Omega_{\rm M}=0.25$, although the fit is less
constraining because of the increased degrees of freedom in the model.
The green curve, instead, depicts what happens with the same freedom
in the model but in the most realistic case when one reconstructs
$\xi(r)$ by deprojecting \xiiz.  Although the estimate of \Om is still
almost unbiased (i.e. the systematic errors are smaller than the
random ones), the minimum is now much shallower, thus indicating lower
constraining power, i.e. a larger statistical uncertainty in the
recovered value.  These errors reflect the uncertainties in the
deprojection procedure, which are responsible for the scatter among
the green curves in Fig.~\ref{fig:xi_sigmaV}.

As previously discussed, the function $\Delta F$ does not obey a
$\chi^2$ statistics and therefore we cannot use the curves in
Fig.~\ref{fig:chi2} to define confidence interval and estimate errors
on \Om.  Ideally, one should repeat the analysis using many halo
catalogues extracted from the simulation.  However, in our analysis we
have already considered the whole computational box and would need to
run more, independent N-body simulations, which are not available.  We
are therefore forced to evaluate errors using techniques that are
typical of error estimates from observational samples. Specifically,
we use the ``block-wise bootstrap'' technique: we divide the box
into 27 independent sub-boxes and build several boostrap samples, each
containing 27 sub-samples selected at random, with replacement, from
the original dataset.  The 1-$\sigma$ errors are then evaluated from
the scatter on the relevant quantities among the bootstrap samples
\citep{norberg2009}.

We apply this technique to quantify the uncertainties on our estimated
values of \Om, $\beta$ and $\sigma_{12}$ when using the procedure
described in this section as applied in a realistic situation,
i.e. redshift-space with free parameters (i.e. the cases of the blue
and green curves in Fig.~\ref{fig:iso_model_OmegaM}).  We obtain a
value \Om=$0.24\pm 0.03$, corresponding to a 1-$\sigma$ uncertainty of
$12\%$, when the {\em true} $\xi(r)$ is used in the RSD model
(i.e. the black curve).  When using the deprojected $\xi(r)$, the
error on \Om grows up to $\sim 40\%$. Still, no systematic bias is
apparent.

Finally, all these results have been obtained assuming no errors on
measured redshifts.  We checked directly that the impact of these
errors on our conclusions is indeed negligible, as long as $\delta
z\lesssim0.2\%$.  However, when $\delta z > 0.3\%$, the resulting
systematic errors on $\beta$ do propagate to \Om and can bias its
estimate.


\subsection {Constraints on curvature and on the DE equation of state}
\label{subsec:AP2}
In this section, we investigate how GD can help in detecting
  possible deviations from a flat $\Lambda $CDM scenario. Let us
  assume a more general DE model with equation of state:
\begin{equation}
\frac{p_{\rm DE}}{\rho_{\rm DE}}=w_0+w_a\frac{z}{1+z} \; ,
\end{equation}
\citep{chevallier2001, linder2003}. In this case, the relevant
cosmological parameters are: \Om, $\Omega_\Lambda$, $w_0$ and $w_a$
(see Eqs.~\ref{eq:Hubble}--\ref{eq:DA}--\ref{eq:xiAP}).

As described in the previous sections, our AP test exploits only
clustering distortions and does not consider the information encoded
in the shape of the correlation function. The advantage is that our
method does not depend on the galaxy bias model. The drawback is that
with no constraints on the shape of $\xi(r)$, the above parameters are
degenerate. This can be seen in Fig~\ref{fig:APpar}, that shows the
$68$ and $95$ per cent pseudo-likelihood probability contours in the
$\Omega_{\rm M}-\Omega_\Lambda$, $\Omega_{\rm M}-w_0$ and $w_0-w_a$
planes. In each plot, the other two parameters that are not shown are
fixed to the true values. The red squares mark the cosmological
parameters of the simulation. These constraints have been obtained in
redshift-space, using the {\em true} $\xi(r)$ and fixing $\beta$ and
$\sigma_{12}$ to their best-fit values as derived with the correct
cosmology, so that the red curve in Fig.~\ref{fig:chi2} corresponds to
the pseudo-likelihood contours along the dotted black line in
Fig~\ref{fig:APpar}, that illustrates the case of a flat universe
($\Omega_\Lambda=1-\Omega_{\rm M}$). The flatness constraint is almost
perpendicular to the degeneracy in the $\Omega_{\rm
  M}-\Omega_\Lambda$, and is the reason that allowed to constrain \Om
in the previous section.


\section{Discussion and Conclusions} 
\label{sec:conclusions}

In this work, we have investigated some relevant limitations existing
when using the anisotropy of galaxy clustering to measure the growth
rate of density fluctuations, while accounting at the same time for
the extra distortions induced by the cosmology-dependent mapping of
redshifts into distances.  More specifically, we have assessed the
impact of different types of uncertainties, both observational and
theoretical, on the estimated values of $\beta$, the anisotropy
parameter closely related to the growth rate. We have then tested how
well, in presence of RSD, the correct underlying cosmology can be
inferred.

The main results of these analyses can be summarized as follows.
\begin{itemize}

\item The impact of Gaussian redshift errors on the estimate of
  RSD can be assimilated to a generalized small-scale Gaussian
  velocity dispersion, which can be quantified in terms of a single
  parameter analogous to the usual pairwise velocity dispersion
  $\sigma_{12}$.

\item In catalogues with volume, density and bias similar to the ones
  analysed in this work, we can estimate $\beta$ from RSD with an
  accuracy of $5-10\%$, regardless of the redshift errors. A general
  scaling formula for the statistical error on $\beta$ as a function
  of the survey parameters is calibrated and presented in the
  companion paper by \citet{bianchi2012}.

\item With typical spectroscopic redshift errors ($\sigma_v \lesssim
  600$ \kms), the anisotropy parameter $\beta$ measured using
  galaxy-sized haloes is systematically underestimated by $\sim
  10\%$. This is discussed in more detail in \citet{bianchi2012},
  where it is also shown that this systematic error depends on the
  bias of the haloes considered.

\item Larger redshift errors ($\sigma_v \gtrsim 1000$ \kms) introduce
  an opposite systematic bias in the estimate of $\beta$, if not
  modelled properly.  This can be partly alleviated using a Gaussian
  model for the velocity distribution function $f(v)$, rather than the
  exponential one.  Note, however, that this may be influenced by the
  fact that a Gaussian distribution has been assumed for redshift
  errors (which is, in any case, a realistic choice for spectroscopic
  observations).

\item A key ingredient in modelling RSD in a sample is a good
  knowledge of the underlying real-space correlation function
  $\xi(r)$.  Random errors on $\beta$ are increased by a factor $\sim
  2$ when $\xi(r)$ is obtained through the deprojection of the
  observed \xiiz, with respect to when using the correct $\xi(r)$.

\item GD arising from an incorrect choice of
  the background cosmology affect both the measured correlation
  function and its model, and thus can impact the estimate of $\beta$.
  However, we have seen that this is very small, meaning that the
  value of $\beta$ can be recovered with similar accuracy even
  assuming a wrong cosmological model.

\item GD have an impact on the estimated galaxy bias. The effect is to
  introduce a spurious scale dependence in the biasing function on
  those scales in which it is supposed to be constant. However, the
  effect is very small and of the same order of theoretical
  uncertainties in current bias models.

\item We have implemented and tested an alternative procedure to
  perform the Alcock-Paczynski test from the observed \xiiz, measuring
  simultaneously $\beta$ and the parameters that enter
  Eq.~(\ref{eq:Hubble}). This is based on the (verified) assumptions
  that the effect of RSD dominates over GD and that the best match
  between RSD observations and the RSD model is realized for the
  correct cosmology.  We have shown that this procedure is robust and
  the results unbiased, in the case of a flat $\Lambda $CDM model.  We
  give a first, approximated estimate of the uncertainty that can be
  expected for \Om through a block-wise bootstrap resampling. In a
  volume $V=2.4\cdot10^9 (h^{-1}\,\mbox{Mpc})^3$, we find that the
  expected errors on \Om are of the order of $\sim12\%$, rising up to
  $\sim40\%$ if the deprojected $\xi(r)$ is used instead of the true
  one.  The results are very insensitive to the accuracy of the model
  used to describe RSD and to the magnitude of redshift measurement
  errors (up to $\delta z\sim0.2\%$). Finally, we have investigated
  how GD can be exploited to constrain both the curvature of the
  Universe and the DE equation of state.
\end{itemize}

In this paper we focused on the analysis of a simulation snapshot
centered at $z=1$.  Clearly, we could have analysed a corresponding
box at $z=0$, but preferred to focus on a redshift range which is
becoming more and more important with ongoing deep surveys like VIPERS
(which has an effective redshift around 0.8 and stretches out to
$z=1.4$ with its brightest galaxies \citep{guzzo2012}), and with
future larger surveys.  Also, we concentrated our analysis on
intermediate scales, $r<50 $\Mpch, where most of the RSD signal lies.
These scales will remain important for these studies also in future
surveys in which larger, even more linear scales will be surely better
sampled, but nevertheless not sufficient alone for reaching the per
cent precisions we are aiming for.

Finally, we also limited our modelling to the simple dispersion model.
We are aware, as we show in our companion paper \citep{bianchi2012},
that this is not a fully appropriate description of clustering and RSD
on such mildly non-linear scales, when the precision on statistical
errors becomes high.  This is very probably at the origin of the
observed $\sim 10\%$ systematic error on the recovered $\beta$ and
significant work is being performed to improve it \citep[see e.g.][and
  references therein]{delatorre2012}. Still, in its simplicity the
dispersion model performs surprisingly well when compared to much more
complicated expressions \citep[e.g.][]{blake2011a} and delivers
statistical errors comparable or smaller than those of more
sophysticated non-linear corrections \citep{{delatorre2012}}.  The
fact that the impact of non-linear effects on estimated errors is
quite limited is also suggested by the close similarity of the errors
on $\beta$ estimated as in this paper to those predicted by a Fisher
matrix analysis \citep{bianchi2012}.


\section*{acknowledgments}
We warmly thank C. Carbone, A. Hawken, A. Heavens and M. Pierleoni for
helpful discussions and suggestions and C. Baugh for supporting this
work through the BASICC simulation. We would also like to thank the
anonymous referee for helping to improve and clarify the paper.  We
acknowledge financial contributions from contracts ASI-INAF
I/023/05/0, ASI-INAF I/088/06/0, ASI I/016/07/0 `COFIS', ASI
`Euclid-DUNE' I/064/08/0, ASI-Uni Bologna-Astronomy Dept. `Euclid-NIS'
I/039/10/0, and PRIN MIUR `Dark energy and cosmology with large galaxy
surveys'.

\bibliographystyle{mn2e} 
\bibliography{bib}

\label{lastpage}

\end{document}